\documentclass[twocolumn]{aastex631}
\usepackage{amsmath, bm, bbm}
\usepackage{amsthm}
\usepackage{amssymb}
\usepackage{mathtools}
\usepackage{physics}
\usepackage{xcolor}
\usepackage{enumitem}
\usepackage{microtype}
\usepackage{graphicx}
\usepackage{booktabs} 
\usepackage{tikz}
\usetikzlibrary{matrix,positioning}
\usetikzlibrary{positioning}
\usetikzlibrary{arrows.meta,calc,decorations.pathreplacing}
\usepackage{listings}
\usepackage{hyperref}
\hypersetup{
    colorlinks,
    linkcolor={red!50!black},
    citecolor={blue!50!black},
    urlcolor={blue!80!blue}
}
\usepackage{url}
\usepackage{needspace}
\usepackage{etoolbox}

\pretocmd{\section}{\Needspace{6\baselineskip}}{}{}
\pretocmd{\subsection}{\Needspace{5\baselineskip}}{}{}
\pretocmd{\subsubsection}{\Needspace{4\baselineskip}}{}{}

\usepackage{listings}
\lstnewenvironment{algolst}[1][]
{
  
  \lstset{#1}
}{}

\DeclareRobustCommand{\bbone}{\text{\usefont{U}{bbold}{m}{n}1}}

\begin{document}

\title{Mapping the Information Geometry of an Unresolved Dark Matter Population using a Differentiable Strong Lensing Simulator}

\author{Alexandre Adam}
\affiliation{Université de Montréal, Montréal, Canada}
\affiliation{Mila - Quebec Artificial Intelligence Institute, Montréal, Canada}
\affiliation{Ciela - Montreal Institute for Astrophysical Data Analysis and Machine Learning, Montréal, Canada}

\begin{abstract}
  Strong gravitational lensing is a unique probe of the matter power spectrum on small scales, 
  where the abundance of dark matter subhalos in galaxies could be used to distinguish 
  the predictions of the concordance cold dark matter model from alternatives such as warm dark matter.
  Extracting this signal poses major computational challenges, 
  since perturbations of the lensing potential induced by substructure can be degenerate with both the macro-model of the lens 
  and highly flexible models of the background source.
  Here, we introduce a framework to quantify these degeneracies using a differentiable strong-lensing simulator.
  Substructure is represented in a spectral basis confined to an annular domain surrounding the lensed image, 
  allowing signals from a population of NFW subhalos to be encoded in a finite vector space.
  We then use the Fisher matrix to determine how much information about substructure is absorbed by nuisance components of the model.
  We find that macro-model degeneracies are largely confined to low-order perturbations of the lensing potential, 
  while degeneracies with the source model can strongly suppress sensitivity across a broad range of scales as the expressivity of the source model is increased.
  Finally, we introduce the Fisher Graph Laplacian prior as a diagnostic tool to study how the internal degeneracies of the source model can be used 
  to regulate the sensitivity of the data to an unresolved population of dark matter subhalos.
\end{abstract}


\section{Introduction}

In a hierarchical structure formation model, galaxies are expected to contain a population of cold dark matter (CDM) subhalos \citep{Springel2017,Bullock2017}.
While individual members of this population can occasionally be detected through their gravitational effects in strong lensing data
\citep{Vegetti2010,Vegetti2012,Hezaveh2016ALMA,Powell2025,Amvrosiadis2026},
the unresolved population contains complementary statistical information about the small-scale distribution of matter.
In particular, the abundance of subhalos at different mass scales can be used to test the predictions of the concordance $\Lambda$CDM cosmology \citep{Planck2020}
against alternatives such as warm dark matter (WDM), where structure formation is suppressed below a characteristic scale \citep{Viel2005,Lovell2014}.

Strong gravitational lensing is one of the few probes capable of accessing this regime.
Current constraints on the subhalo mass function have been obtained from the frequency of individual detections \citep{Vegetti2014} 
and from the amplitude of flux anomalies in lensed quasars \citep{Gilman2020,Gilman2026}.
Beyond these approaches, 
an unresolved population of low-mass subhalos should collectively produce a stochastic imprint on the lensed images of extended background galaxies.
This signal could be used to constrain the small-scale, or low-mass, content of the matter power spectrum \citep{Hezaveh2016PS,DiazRiveiro2018,CyrRacine2019}.
However, despite observations with sufficient angular resolution to carry out this program \citep[e.g.][]{Bolton2006,Bolton2008},
precise constraints on the small-scale sector remain out of reach.

The most promising route to this signal is to extract information from highly magnified arcs, 
where small perturbations of the lens mapping can produce anomalies in the data \citep[for a review, see][]{Vegetti2024}.
In this regime, the morphology of the background source acts as a precision instrument to probe the gravitational potential.
However, the true unlensed image of the source is fundamentally unknown.
A model for the source and a model for the foreground mass distribution must therefore be inferred jointly from the same data.
This inversion is plagued by degeneracies, where small adjustments of the source model can mimic distortions that would otherwise 
be attributed to perturbations of the lensing potential \citep{Nightingale2024,Ephremidze2025}.

Overcoming this structural difficulty presents a steep challenge when confronting the subtle signals expected from a population of low-mass subhalos.
To access this information, highly flexible source models must be deployed to reduce systematic biases \citep{Legin2025},
thus exacerbating existing degeneracies with substructure. 
Although exploratory works have begun studying these interactions \citep{Vernardos2022}, 
a rigorous framework for quantifying degeneracies in high-dimensional spaces is still lacking.

In this work, we build such a framework on top of the open-source strong-lensing simulator \texttt{Caustics} \citep{Stone2024}, natively implemented in \texttt{PyTorch} \citep{torch}, allowing us to take advantage of a fully differentiable simulator.
Given fiducial parameters, degeneracies can be quantified in a locally Gaussian approximation of the Bayesian posterior
where the information geometry of the parameters is encoded by the Fisher matrix.
To maintain computational tractability in high-dimensional settings, we make use of matrix-vector products computed by automatic differentiation, 
allowing us to keep track of the information geometry even when the Fisher matrix cannot be materialized explicitly.
This construction provides a blueprint for tracking how information about substructure is absorbed by high-dimensional nuisance models during marginalization.

Within this framework, we introduce two core technical developments.
First, we represent substructure as perturbations of the lensing potential in a spectral basis defined on an annulus surrounding highly magnified images.
This basis provides a finite vector space for the collective signal of an unresolved population of subhalos, 
while matching the geometry of the Einstein ring.
The implementation of this basis is differentiable, enabling us to capture the degeneracies between substructure and nuisance parameters in the simulator. 
The choice of an annulus geometry is inspired by gravitational imaging methods \citep{Koopmans2005} and the recently introduced curved arcs basis \citep{Sengul2025}.

Second, we introduce the Fisher Graph Laplacian (FGL) prior as a diagnostic tool for studying the role of source regularization.
The FGL prior constructs a graph from the off-diagonal entries of the source Fisher matrix, 
thereby using the internal degeneracies of the source model to regulate its expressivity.
By varying the strength of this prior, we can test how much of the inferred dark-matter sensitivity is conditional on this regularization.

Our results show that the macro-model and source model affect substructure information in qualitatively different ways.
The macro-model primarily absorbs large-scale perturbations of the lensing potential, leaving most high-frequency substructure modes nearly unchanged.
By contrast, the source model can absorb substructure responses across a much broader range of scales, 
with the effect becoming stronger as the source resolution is increased.
The amount of information that survives marginalization is therefore controlled primarily by the expressivity of the nuisance model 
used to interpret the data.

The remainder of this paper is structured as follows.
In Section~\ref{sec:tangent_space}, we describe the information geometry formalism used to quantify degeneracies.
In Section~\ref{sec:substructure_potential}, we present the annulus eigenbasis used to represent substructure.
Section~\ref{sec:methods} details the computational pipeline, including the FGL prior.
Results are shown in Section~\ref{sec:results} and discussed in Section~\ref{sec:discussion}.
We conclude in Section~\ref{sec:conclusion}.

\section{The tangent space of a differentiable simulator}\label{sec:tangent_space}

Broadly speaking, the search for dark matter substructure in strong gravitational lensing data can be framed as a Bayesian inference problem. 
Given data $\bm{d}$, the goal is to sample from the posterior distribution over the parameters of interest $\vartheta$, describing substructure,
while marginalizing over nuisance parameters $\eta$ describing confounding factors in the data.
Concretely, the posterior can be written in terms of the likelihood and prior using Bayes' theorem
\begin{equation}
  p(\vartheta \mid \bm{d}) \propto \int p(\bm{d} \mid \vartheta, \eta) p(\vartheta, \eta) \, \dd \eta\, .
\end{equation} 

In this section, we study this posterior using a locally Gaussian approximation centered on a fiducial simulator model.
By taking the linear expansion of a differentiable simulator around the fiducial parameters,
we can translate Bayesian marginalization into the language of information geometry \citep[for a review, see][]{Amari2016,Nielsen2018},
where it becomes analogous to a projection against the tangent space of the nuisance parameters.
Crucially, this language provides a rigorous definition for the degeneracy between parameters, 
which is the central focus of this work.

\subsection{Linearized forward model}

\begin{figure}[t!]
\centering
\includegraphics[height=5.5cm]{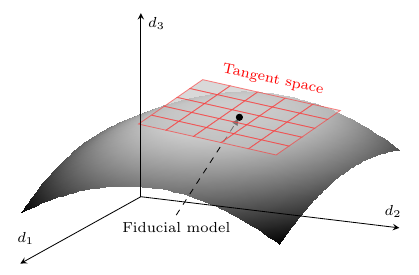}
\caption{Schematic representation of the simulator manifold, a curved surface in data space, and its tangent space.
}
\label{fig:manifold}
\end{figure}

To analyze gravitational lensing data, a simulator $\bm{f}(\theta)$ is designed with a set of parameters that describes the lensed image.
The output of the simulator is related to the data by the forward model
\begin{equation}
  \bm{d} = \bm{f}(\theta) + \bm{n}\, ,
\end{equation} 
which includes a source of additive noise that accounts for random measurement errors.
For simplicity, we ignore structured source of noise, 
which permits us to approximate its distribution with a Gaussian
\begin{equation}
  \bm{n} \sim \mathcal{N}(0, C)\, ,
\end{equation} 
where $C$ is a positive semidefinite covariance matrix. 
To study the information geometry of the data, we typically rely on the precision matrix $C^{-1}$, 
which corresponds to the metric used to compare data vectors and estimate their norm, 
i.e.~$\left\lVert \bm{u} \right\rVert^{2}_{C} = \bm{u}^{\top}C^{-1}\bm{u}$.

Since the effects of substructure are assumed to be weak \citep[e.g.][]{Dalal2002,Hezaveh2016PS},
the natural regime to study the problem is a linear expansion of the simulator around fiducial parameters
\begin{equation}
  \bm{f}(\theta_0 + \Delta \theta) \simeq f(\theta_0) + \frac{\partial \bm{f}}{\partial \theta}\bigg|_{\theta_0} \Delta \theta\, .
\end{equation} 
The columns of the Jacobian matrix 
\begin{equation}
  J_i \equiv \frac{\partial \bm{f}}{\partial \theta_i}\bigg|_{\theta_0}\, ,
\end{equation} 
are data space vectors that represent the response of the simulator to small perturbations in the parameters.
The span of its columns defines the tangent space of the simulator manifold (see Figure \ref{fig:manifold}).

In what follows, we refer to an anomaly as a column vector from the substructure tangent space.
Whether we can identify such an anomaly in the data depends on its similarity with the responses of nuisance parameters. 
To make this explicit, we split the parameters into quantities of interest and nuisance parameters $\theta = ( \vartheta,\, \eta)$.
Accordingly, we split the Jacobian matrix into blocks of columns $J = (J_{\vartheta},\, J_{\eta})$.

\subsection{The Fisher matrix}

To quantify degeneracies, we compare parameter responses using the data-space metric.
Doing so for every pairs of responses defines the Fisher matrix
\begin{equation}
  F = J^{\top} C^{-1} J = 
  \begin{pmatrix}
    F_{\vartheta\vartheta} & F_{\vartheta \eta} \\
    F_{\eta \vartheta} & F_{\eta \eta}
  \end{pmatrix}\, .
\end{equation} 
In a Bayesian context, $F$ is the inverse covariance of the likelihood. 
Here, it is understood as a metric specifying the information geometry of parameter space.
In any case, its diagonal elements measure the local information that each parameter carries about the data,
while off-diagonal components accounts for degeneracies between them.
Two parameters are degenerate when their responses are nearly colinear,
while orthogonal responses allow us to perfectly distinguish the parameters from the data.

\subsection{The geometry of marginalization}\label{sec:geometry}

If an anomaly can be produced by changing the nuisance parameters, then it cannot be uniquely attributed to substructure.
In the locally Gaussian posterior approximation, 
the nuisance model is linear and the anomalies it absorbs lie in the tangent space spanned by the columns of $J_{\eta}$.

This statement can be written as the solution to a least-squares problem where we seek the nuisance parameters 
$\bm{x}$ whose response $J_{\eta}\bm{x}$ best matches the anomaly $\bm{b}$.
If the nuisance parameters are regularized by a Gaussian prior with precision $\Lambda_{\eta}$, 
then the maximum a posteriori solution can be written as 
\begin{equation}
\bm{x}^{\star} = \underset{\bm{x}}{\mathrm{arg\,min}}\, \left[ \left\lVert \bm{b} - J_{\eta}\bm{x} \right\rVert^{2}_{C} + \bm{x}^{\top}\Lambda_{\eta}\bm{x} \right]\, .
\end{equation}
By taking derivatives of this objective function and setting them to zero, we obtain the normal equation
\begin{equation}
G_{\eta}\bm{x}^{\star} = J_{\eta}^{\top}C^{-1}\bm{b}\, ,
\end{equation}
where $G_{\eta} = F_{\eta \eta} + \Lambda_{\eta}$ is the posterior metric.

The component of the anomaly absorbed by the nuisance model is obtained by
\begin{equation}
  P_{\eta}\bm{b} = J_{\eta}\bm{x}^{\star}\, ,
\end{equation}
where $P_{\eta} = J_{\eta}G_{\eta}^{-1}J_{\eta}^{\top}C^{-1}$ is analogous to a projection operator onto the nuisance tangent space.
When the prior is uninformative, i.e. $\Lambda_{\eta}=0$, the projection is exact. 
More generally, a prior will penalize some nuisance perturbations, thus reducing the amount of anomalies absorbed by them.

\begin{figure}[t!]
  \centering
  \includegraphics[height=4cm]{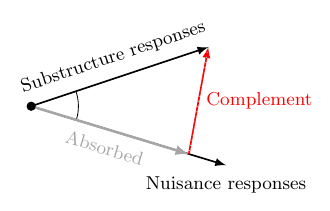}
  \caption{Schematic representation of the substructure and nuisance responses.
    Absorbed anomalies are tangent to the nuisance responses
    while anomalies uniquely attributed to substructure belong to the complement space.
    In general, imposing a prior on nuisance parameters changes how much of an anomaly can be absorbed by the nuisance model.
  }
  \label{fig:projection}
\end{figure}

The component of the anomaly that remains after marginalization is obtained with the complement operator
\begin{equation}\label{eq:complement}
  P_{\eta}^{\perp} = \bbone - P_{\eta}\, .
\end{equation}
Thus, marginalization becomes equivalent to decomposing an anomaly into an absorbed component and its complement,
which remains identifiable in the data (see Figure~\ref{fig:projection}).

\subsection{The transfer function}\label{sec:marginal}

The information that remains after marginalization is obtained by taking the 
Schur complement of the joint posterior metric
\begin{equation}
  F^{\perp}_{\vartheta} = F_{\vartheta \vartheta} - F_{\vartheta \eta} G_{\eta}^{-1} F_{\eta \vartheta}\, .
\end{equation} 
Alternatively, we can write the marginal information using the complement operator 
$F^{\perp}_{\vartheta} = J_{\vartheta}^{\top} C^{-1} P^{\perp}_{\eta} J_{\vartheta}$,
such that obtaining the marginal information amounts to projecting all columns of the substructure Jacobian, 
as described in Section~\ref{sec:geometry}

To quantify how much information is lost to the nuisance model, 
we define the transfer function for an individual substructure component $a$ as the ratio
\begin{equation}\label{eq:transfer_function}
  T_a = \left( \frac{F^{\perp}_{aa}}{F_{aa}} \right)^{1/2}\, .
\end{equation} 
In practice, $F_{aa}^{-1/2}$ corresponds to a nominal error on mode $a$ if the nuisance model was known perfectly,
whereas $(F^\perp_{aa})^{-1/2}$ is the marginalized error. 
The transfer function is thus an efficiency factor, 
where $T_a \simeq 1$ is associated with an anomaly that remains identifiable in the data 
and $T_a \simeq 0$ signifies that the anomaly is completely degenerate with nuisance responses.

In the results, we use $T_a$ as a diagnostic to determine which substructure modes survive
marginalization over the macro-model and background source model.

\section{The substructure potential}\label{sec:substructure_potential}

In the thin-lens approximation, the theory of strong gravitational lensing is governed by a potential function \citep{Blandford1986,Schneider1992}.
This potential is sensitive to localized perturbations,
such that the observable signatures of dark matter are confined to a narrow region near the lensed image. 
In this section, we introduce a set of spectral coordinates for such perturbations.

\subsection{The lensing potential}

To simulate the lensed image, we use the lens equation
\begin{equation}
  \bm{\beta}(\bm{\theta}) = \bm{\theta} - \bm{\alpha}(\bm{\theta})\, ,
\end{equation} 
which relates angular coordinates of light rays in the source plane ($\bm{\beta}$) to the image plane ($\bm{\theta}$).
In the regime relevant for substructure, the reduced deflection angles can be related to the gradient of 
the lensing potential $\bm{\alpha}(\bm{\theta}) = \grad \psi(\bm{\theta})$.
For simplicity, we omit line-of-sight effects \citep[for their treatments, see e.g.][]{Sengul2020,Fleury2021}.

The lensing matrix is the Jacobian of the lens equation 
\begin{equation}
  A(\bm{\theta})
  \equiv \frac{\partial \bm{\beta}}{\partial \bm{\theta}} = 
  \bbone - \grad\grad\psi(\bm{\theta}) .
\end{equation}
The magnification of source-plane areas to the image plane is determined by the determinant of the lensing matrix.
In strongly lensed systems, the magnification becomes singular on a locus of points called the critical curve, where $\mathrm{det} A = 0$.
The caustic is the critical curve mapped to the source plane.

The convergence is related to the lensing potential by Poisson's equation
\begin{equation}
  \grad^{2} \psi(\bm{\theta}) = 2 \kappa(\bm{\theta})\, .
\end{equation} 
We decompose the convergence into a smooth macro component --- typically associated with the dominant lensing galaxy 
and its environment --- and a perturbative component associated with substructure.

\subsection{Localizing sensitivity to substructure}

\begin{figure}[t]
  \centering
  \begin{tikzpicture}
    \node at (-4.5, 2.15) {Source};
    \node at (-4.5, 0) {\includegraphics[width=4cm]{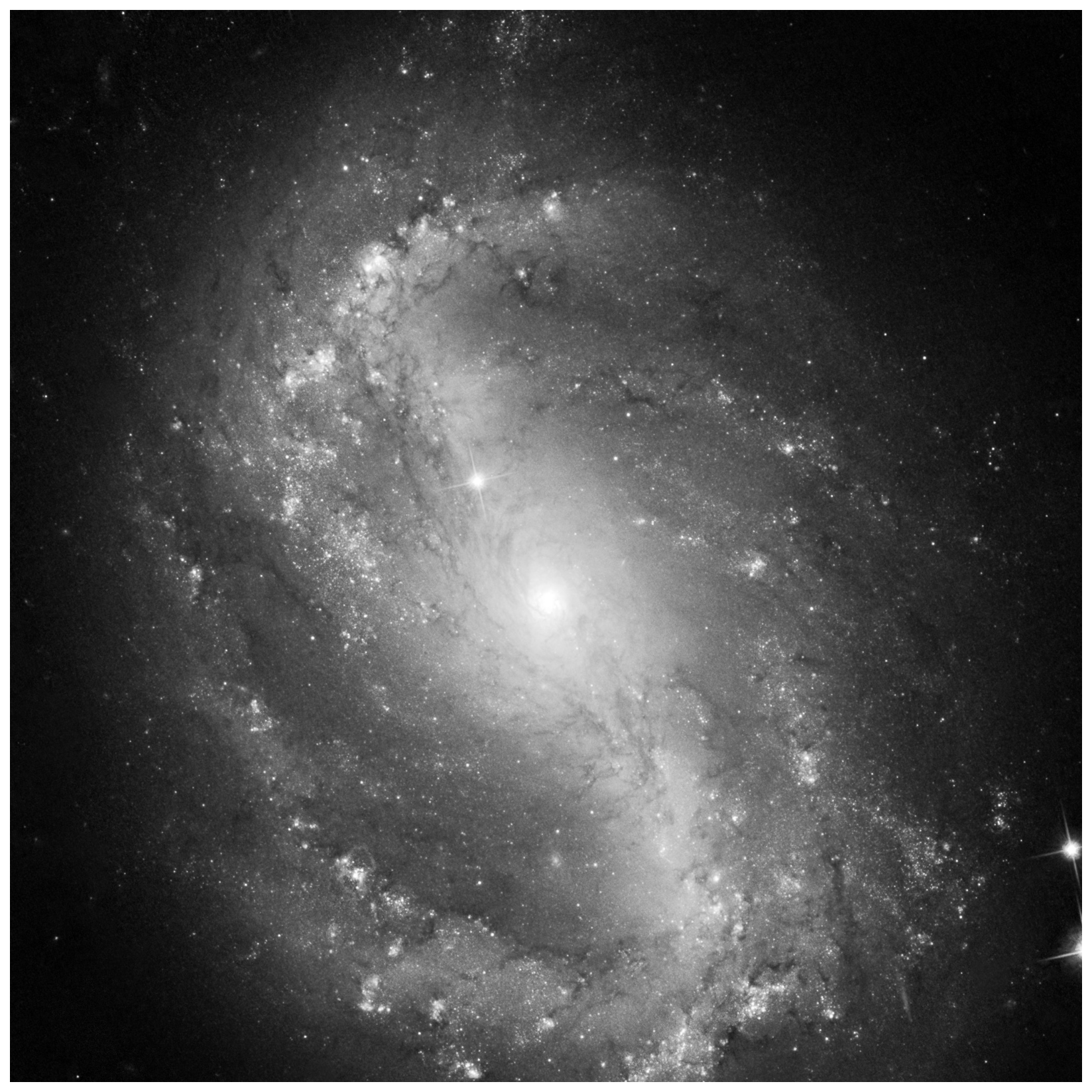}};
    \node at (0, 2.15) {Lensed image};
    \node at (0, 0) {\includegraphics[width=4cm]{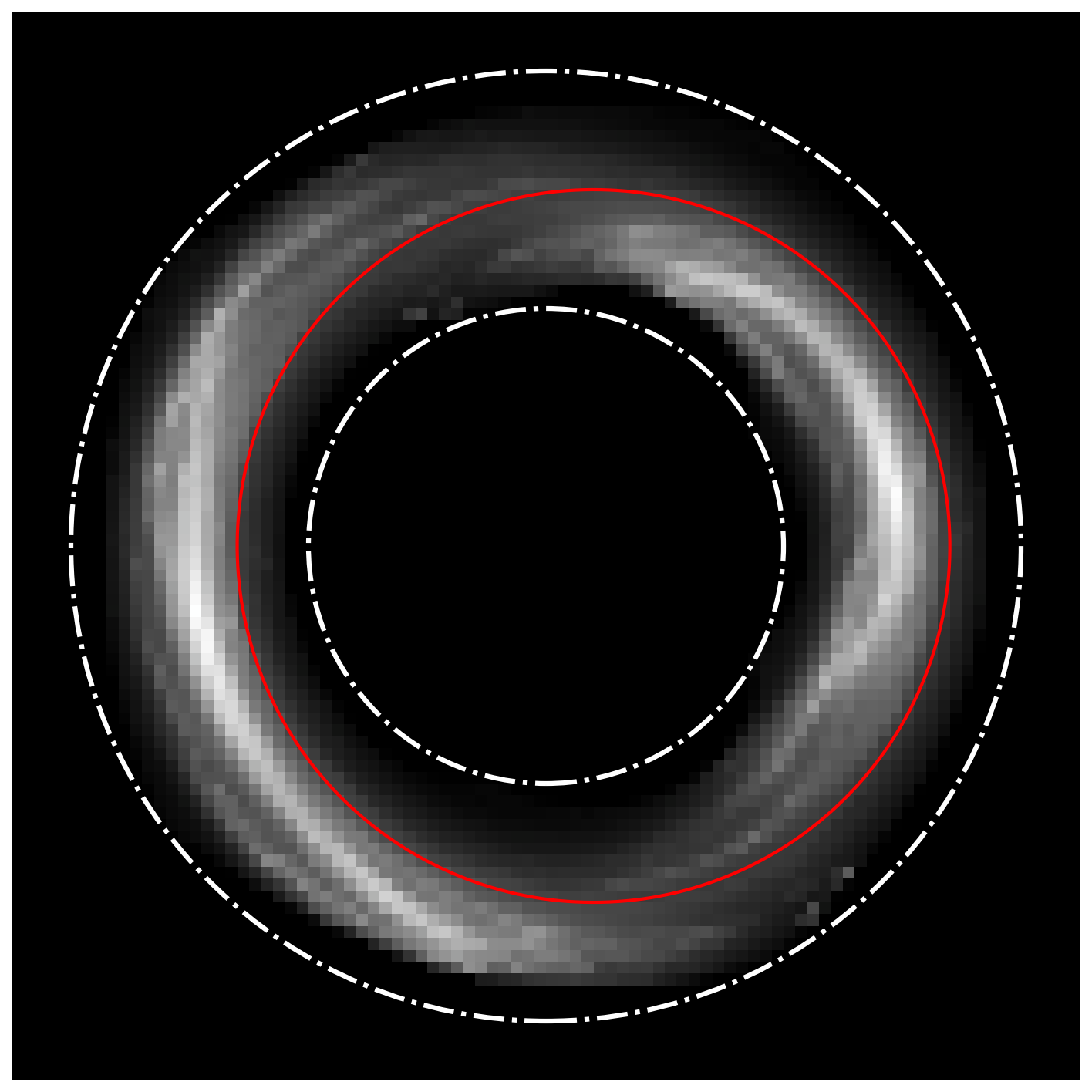}};
    \node[color=red, anchor=west] (cc) at (-2.07, 1.8) {\scriptsize \textbf{Critical}};
    \node[color=red, anchor=west] (cc) at (-1.95, 1.5) {\scriptsize \textbf{curve}};
    \draw[red, -latex, line width=0.4pt]
      (cc) -- (-1, 0.7);
    \node[color=white, anchor=south] (annulus) at (1.3, -2) {\scriptsize \textbf{Annulus}};
    \draw[white, -latex, line width=0.4pt]
      (annulus.north) -- ++(-0.15, .35);
  \end{tikzpicture}
  \caption{
    The left panel shows an HST/ACS image of the barred spiral galaxy NGC~6217 
    (created as part of proposal 11371) 
    used as the fiducial source brightness distribution for the simulator.
    The right panel shows the lensed image corresponding to the \textit{arcs} configuration,
    where the red curve marks the tangential critical curve
    and the white annulus indicates the region where substructure perturbations are defined.
  }
  \label{fig:lensed_image}
\end{figure}
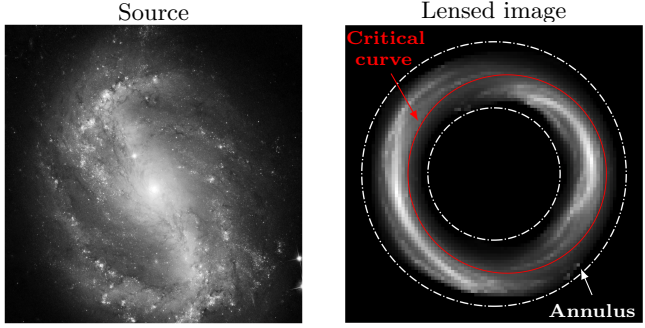

To identify the regions where substructure produces an observable signal, it is useful 
to write the image formation model as
\begin{equation}
  I(\bm{\theta}) = S(\bm{\beta}(\bm{\theta}))\, ,
\end{equation} 
where $I$ denotes an idealized lensed image before it is measured by the detector, 
and $S$ denotes the source brightness distribution.
Schematically, a small perturbation to the lensing potential displaces the 
coordinate of light rays, producing image residuals \citep[e.g.][]{Vegetti2009}
\begin{equation}\label{eq:residuals}
  \delta I = - \grad S \cdot \grad \delta \psi\, .
\end{equation}
Such residuals appear in the data when substructure deflects light rays across resolved source gradients in the source plane.

For typical galaxy-scale lenses, highly-magnified arcs can be found near the tangential critical curve,
where the lens mapping stretches the source and renders visible its small-scale structures over extended regions in the data, 
as in Figure \ref{fig:lensed_image}. 
This gives a motivation for describing potential perturbations supported on an annular domain.

\subsection{Spectral coordinates on the annulus}

A spectral basis corresponds to the set of eigenfunctions $\{\phi_a\}$ that diagonalize the Laplacian
\begin{equation}
  \grad^{2} \phi_{a}(\bm{\theta}) = - k_{a}^{2} \phi_{a}(\bm{\theta})\, .
\end{equation} 
This basis organizes perturbations to the lensing potential into a hierarchy of scales,
which is particularly well suited to describe an unresolved population of dark matter halos
associated with a stochastic fluctuation in the convergence \citep{Hezaveh2016PS}.

The rotational symmetry of the annulus domain allows each eigenfunction to be decomposed 
into an azimuthal Fourier mode and a radial component described by Bessel functions.
We impose Neumann boundary conditions at the inner and outer radii of the annulus, 
so that nonconstant modes have zero total mass and only redistribute the convergence 
within the annulus. The explicit construction is given in Appendix~\ref{sec:neumann_bc}.

For an arbitrary substructure realization supported on the annulus $\bm{\theta} \in \mathcal{A}$,
the convergence can be expanded in the eigenbasis and summarized in terms of a set of coefficients $\{c_a\}$.
The deflection angles 
\begin{equation}
  \bm{\alpha}_{\rm sub}(\bm{\theta})
  =
  -2 \sum_a k_a^{-2} c_a \grad \phi_a(\bm{\theta})\, ,
\end{equation}
can then be added to the lens equation to simulate its effects.

\subsection{Finite-resolution effects}

\begin{figure}[t]
  \centering
  \begin{tikzpicture}
    \node at (2.7, -1.8) {\textbf{a)} Substructure modes};
    \node at (2.7, -5.3) {\textbf{b)}  Responses};
    \node at (0, -0.2) {\includegraphics[width=3cm]{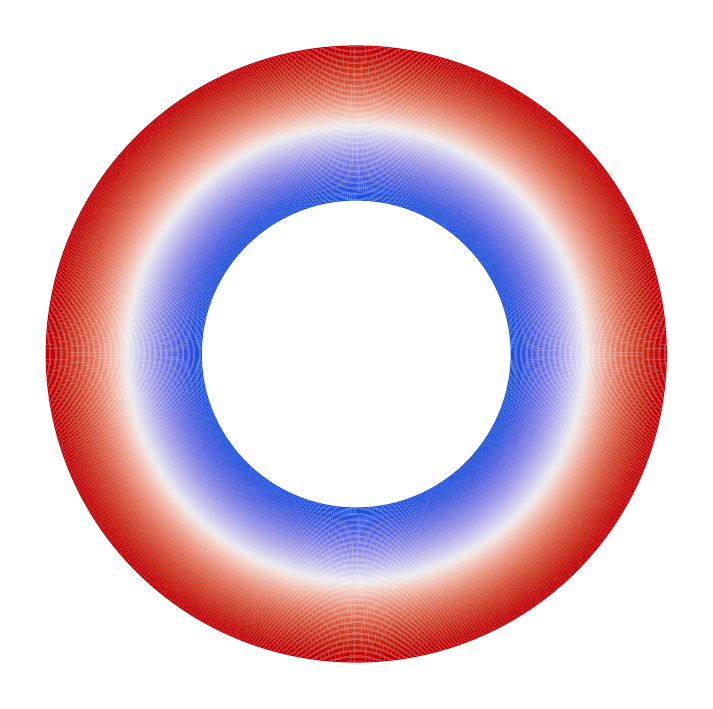}};
    \node at (2.7, 0) {\includegraphics[width=3cm]{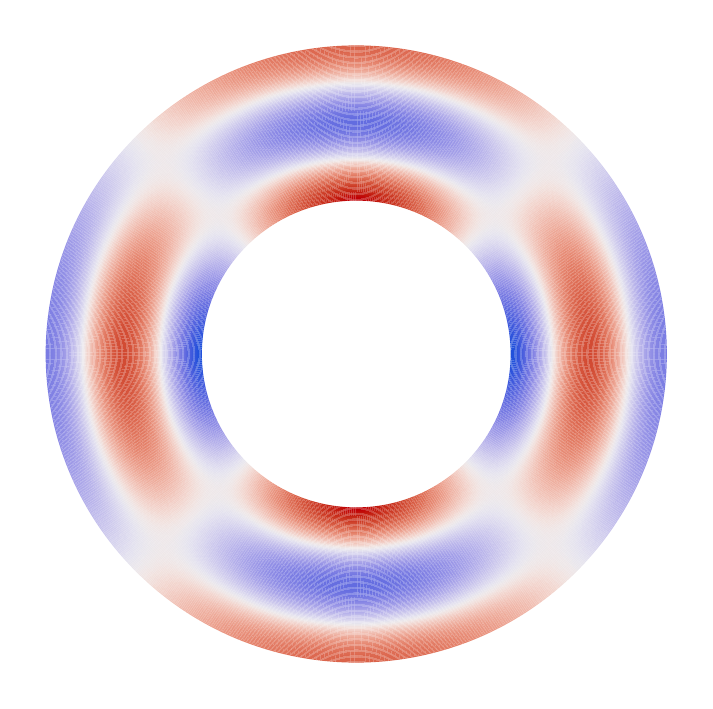}};
    \node at (5.4, 0) {\includegraphics[width=3cm]{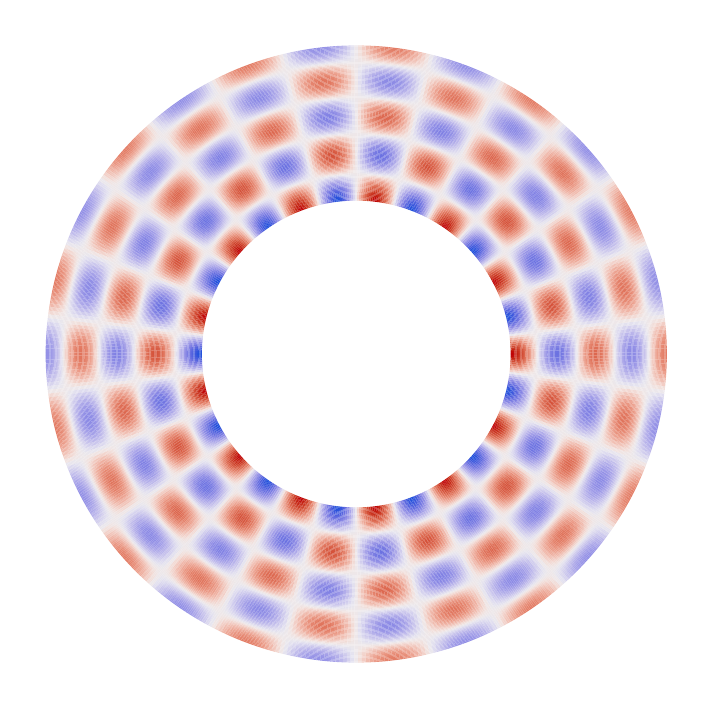}};
    \node at (0, -3.5) {\includegraphics[width=3cm]{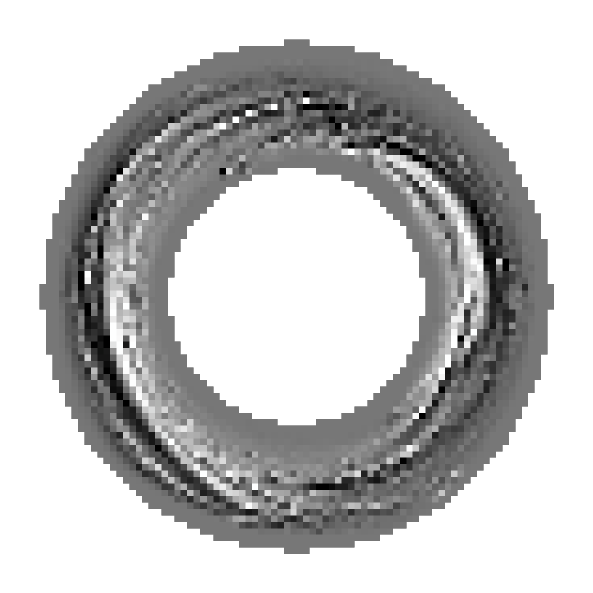}};
    \node at (2.7, -3.5) {\includegraphics[width=3cm]{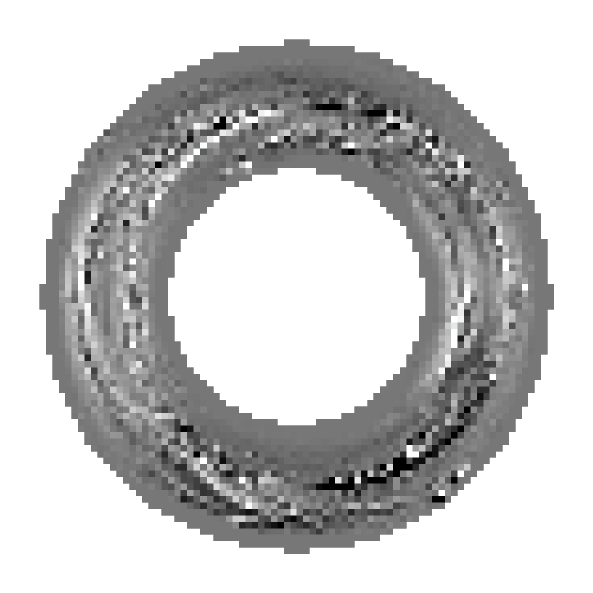}};
    \node at (5.4, -3.5) {\includegraphics[width=3cm]{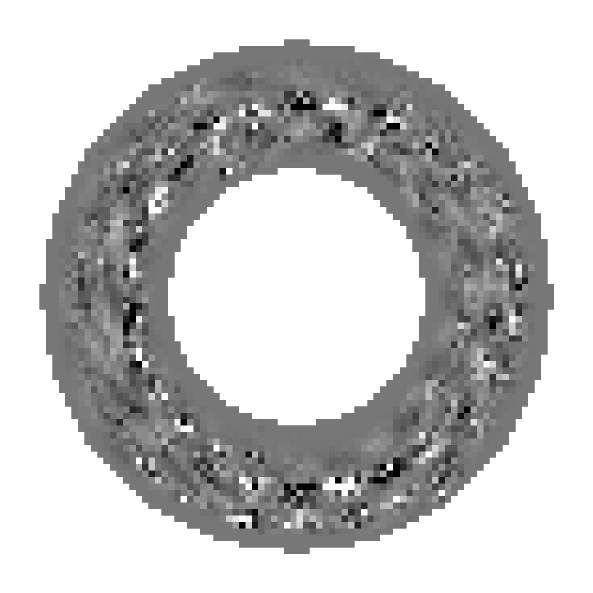}};
  \end{tikzpicture}
  \caption{Examples of \textbf{a)} eigenfunctions of the annulus eigenbasis and \textbf{b)} their response in data space. 
    The blue/red colormap represents negative/positive convergence contrasts, 
    while the gray scale shows residuals.
  }
  \label{fig:modes}
\end{figure}

Since an observation only contains a finite number of resolved degrees of freedom,
the relevant coefficient space is a finite band of modes whose induced image-plane responses vary on scales that can be distinguished by the data.
The effective resolution of the data defines the Nyquist wavenumber $k_{\mathrm{N}}$,
which we use as a guide to truncate the eigenbasis to modes with $k_a \lesssim k_{\mathrm{N}}$.
This cutoff is not a detectability threshold, however,
since the characteristic scale of an image-plane response ultimately depends on the simulator (see Figure \ref{fig:modes}).

\subsection{Dark matter signals in coefficient space}\label{sec:inner_product}

To study the effect of an unresolved population of dark matter subhalos,
we generate NFW subhalos from two mass functions associated with CDM and WDM populations (see Appendix~\ref{sec:population_model}).
The coefficients for one such realization can be computed with an inner product
\begin{equation}
  c_a
  = 
  \int_{\mathcal{A}} \kappa_{\mathrm{sub}}(\bm{\theta}) \phi_a(\bm{\theta}) \dd \bm{\theta}\, ,
\end{equation} 
such that the distribution of coefficients generated by sampling many realizations of $\kappa_{\mathrm{sub}}$ can be used to make theory predictions.

\section{Methods}\label{sec:methods}

In this section, we describe the simulator pipeline and the Fisher Graph Laplacian prior.
The framework is implemented in a custom software package built natively in \texttt{PyTorch} \citep{torch}.
Many of the strong lensing simulator components were built upon the open-source package \texttt{Caustics} \citep{Stone2024}.
All implementation details and experiment configurations ---
including the differentiable annular basis for substructure --- 
will be made publicly available upon publication.
Throughout, we assume the standard \textit{Planck} $\Lambda$CDM cosmology \citep{Planck2020}.

\subsection{Detector model}

We adopt an image-plane pixel scale of $\Delta \theta = 0.05''$, 
comparable to the sampling scale of \textit{Hubble Space Telescope} (\textit{HST}) imaging, 
and define the corresponding Nyquist scale
\begin{equation}
  k_{\mathrm{N}} = \frac{\pi}{\Delta \theta}\, ,
\end{equation} 
as a nominal reference for the substructure mode frequencies that are measurable in the data.
We simulate lenses at redshift $z_{\ell} = 0.5$ with an Einstein radius $R_{\rm E}=1.5''$. 
The field of view of the image plane is set to $4.5''$, which corresponds to 90 data pixels on a side, 
and the substructure annulus is taken to span $r_{\rm in}=1''$ to $r_{\rm out}=2''$. 
For this geometry, the annulus contains 3680 image pixels, which we use as a mask to reduce the number of light rays 
needed to simulate the lensed image.
The source is represented on a square field of view with a width of $6\,\mathrm{kpc}$ at the source redshift $z_s = 1.0$.

To isolate the interplay of substructure with the macro-model and the source model, 
we intentionally simplify the detector response to be uniform within each data pixel.
In particular, we do not convolve the lensed-image with a point-spread function in the simulator. 

\subsection{Macro-model}\label{sec:macro}
We employ an Elliptical Power-Law (EPL) convergence profile \citep[e.g.][]{Tessore2015} for the dominant lensing galaxy
\begin{equation}
  \kappa_{\mathrm{EPL}}(\theta_1, \theta_2) = 
  \frac{2 - t}{2}\left(\frac{R_{\mathrm{E}} \sqrt{q}}{\sqrt{q^{2}\theta^{2}_1 + \theta^{2}_2}} \right)^{t}\, ,
\end{equation}
where $(\theta_1, \theta_2)$ are the coordinates of the principal axis of an ellipse.
The coordinates are shifted horizontally by $x_0$ compared to the origin of the lens plane.
The other parameters are the vertical shift $y_0 = 0$, the Einstein radius $R_{\mathrm{E}} = 1.5''$, axis ratio $q$ and logarithmic slope $t=1$, 
corresponding to an isothermal singular profile.
The parameters that vary between configurations are enumerated in Table~\ref{tab:lens_configurations}.

\begin{table}[htb!]
\centering
\caption{
Lens configurations.
}
\label{tab:lens_configurations}
\begin{tabular}{lcc}
\hline
Configuration & $q$ & $x_0$  \\
\hline
\textit{ring} & 1.0 & 0.0 \\
\textit{double} & 1.0 & 0.5  \\
\textit{quad} & 0.7 & 0.3  \\
\textit{arcs} & 1.0  & 0.2 \\
\hline
\end{tabular}
\end{table}

To explore departures from an elliptical mass profile in the dominant lensing galaxy,
we include multipole moments 
\begin{equation}
  \kappa_{\ell}(\theta_1, \theta_2) = \frac{a_{\ell}}{2 r} \cos \big( \ell(\varphi - \varphi_\ell) \big)\, ,
\end{equation}
where $r=\sqrt{\theta_1^{2} + \theta_2^{2}}$, $\varphi = \mathrm{atan2}(\theta_2, \theta_1)$ and $\ell \in \{3, 4, 5\}$.
Each multipole component is parametrized by an amplitude $a_\ell$ and orientation $\varphi_\ell$.

Finally, external shear is included as an additional deflection field 
\begin{equation}
  \bm{\alpha}_{\mathrm{ext}}(\theta_1, \theta_2) = \gamma_{\mathrm{ext}}
  \begin{bmatrix}
    \theta_1 \cos (2 \varphi_{\mathrm{ext}} )
    + \theta_2 \sin (2 \varphi_{\mathrm{ext}}) \\
    \theta_1 \sin (2 \varphi_{\mathrm{ext}}) 
    - \theta_2 \cos (2 \varphi_{\mathrm{ext}})
  \end{bmatrix}\, ,
\end{equation}
parametrized by an amplitude $\gamma_{\mathrm{ext}}$ and orientation $\varphi_{\mathrm{ext}}$.
For simplicity, the effect of multipoles and external shear are studied in the vicinity of vanishing amplitudes ${a_{\ell} = \gamma_{\mathrm{ext}} = 0}$ 
and their orientation is aligned with the horizontal axis.

In total, the macro-model is composed of 14 parameters.
The tangent-space analysis includes perturbations of the macro-model around the fiducial values enumerated in this section.

\subsection{Substructure Model}\label{sec:substructure_model}

Substructure is represented by a finite set of annulus coefficients $\{c_{a}\}_{a=1}^{N}$, 
where the index $a = (m, n, p)$ is composed of the azimuthal order $m \in \{0, \ldots, m_{\mathrm{max}}\}$, 
the radial order $n \in \{1,\ldots, n_{\mathrm{max}}\}$ and angular parity of the mode $p \in \{\cos, \sin\}$.
The truncations $n_{\mathrm{max}}$ and $m_{\mathrm{max}}$ are based on the effective resolution of the data.

In practice, the wavenumbers of the annulus eigenfunctions are computed by solving for the roots of a transcendental equation (see Appendix~\ref{sec:roots}).
To expedite the search for the truncations, we use a WKB approximation for the radial Bessel functions (see Appendix~\ref{sec:wkb}).
As a result, we can estimate the truncations from the geometry of the annulus
\begin{equation}
  m_{\rm max}
  \sim
  \frac{\pi r_{\rm out}}{\Delta\theta},
  \qquad
  n_{\rm max}
  \sim
  \frac{r_{\rm out}-r_{\rm in}}{\Delta\theta} .
\end{equation}
We use the nearest integer greater than this estimate, 
corresponding to $m_{\mathrm{max}} = 126$ and $n_{\mathrm{max}} = 20$, 
or $N=5060$ modes in total when including the parity indices.

For numerical efficiency, Bessel functions and their derivatives are precomputed on an equally spaced grid of knots.
The radial components entering the deflection angles are then evaluated using linear interpolation, which is fast and differentiable.
To accurately resolve the smallest radial oscillations, we adopt a grid density of $5\,n_{\rm max}$ knots.

To connect substructure perturbations to a physical dark matter population, 
we generate explicit realizations of subhalos and compute their coefficients using the inner product defined in Section~\ref{sec:inner_product}.
Each subhalo is modeled with an NFW profile \citep{Navarro1996,Navarro1997}, 
parametrized by the enclosed mass, concentration and position $\theta_h$.
The convergence is defined as
\begin{equation}
  \kappa_{h}(\bm{\theta})
  =
  \kappa_s F_{\mathrm{NFW}} \left( \frac{\bm{\theta} - \bm{\theta}_h}{\theta_s}  \right)\, ,
\end{equation}
where $\theta_s$ is the scale radius, 
$\kappa_s$ is a normalization determined by the enclosed mass and redshift of the lens,
and $F_{\mathrm{NFW}}$ is the projected NFW kernel \citep{Bartelmann1996}.
The mass-concentration relation and population sampling procedure are described in Appendix~\ref{sec:population_model}.

\subsection{Source model}\label{sec:source_model}

The unlensed image of the source is discretized with a Cartesian grid of $N_s \times N_s$ pixels.
Throughout our experiments, the field of view is maintained fixed, such that a larger number of pixels directly 
corresponds to increasing the expressivity of the model.
For our experiments, we consider $N_s \in \{64,128,256,512,1024,2048\}$.
The upper bound of this range is chosen to approximate a continuum limit, 
where the source resolution has become fine enough to resolve the majority of highly magnified details in the data,
given the macro-model and detector configuration used throughout this work.

\subsection{Adaptive ray tracing}\label{sec:adaptive}

A detector pixel measures the flux integrated over a finite image-plane area.
Through the lens mapping, this pixel has an associated source-plane collecting area 
where the surface-brightness of the source model is averaged to simulate the integrated flux.
Inspired by the work of \citet{Karchev2022}, we represent the collecting area by a kernel $K_q$ normalized to unity.
The integrated flux of a detector pixel $i$ then becomes 
\begin{equation}
  d_i = \sum_{q} w_q \int S(\bm{\beta}) K_q(\bm{\beta}) \, \dd \bm{\beta}\, ,
\end{equation} 
where $w_q$ are quadrature weights.

For a sufficiently small quadrature element located at $\bm{\theta}_q$,
we can approximate the kernel using a Gaussian distribution
\begin{equation}
  K_q(\bm{\beta}) \approx \mathcal{N}(\bm{\beta} \mid \bm{\beta}(\bm{\theta}_q),\, \Sigma_{\beta}(\bm{\theta}_q) + \Sigma_{s})\, ,
\end{equation}
where $\mathcal{N}(\cdot \mid \mu, \Sigma)$ denotes the normal distribution with mean $\mu$ and covariance matrix $\Sigma$.
Crucially, $\Sigma_{\beta}(\bm{\theta}_q)$ encodes an image-plane area pushed-forward to the source-plane by the lens equation.
An elementary derivation for this covariance matrix, as well as $\Sigma_s$, is given in Appendix~\ref{sec:adaptive_kernel}.
To reduce the computational burden of ray tracing, the kernels are truncated at three standard deviations before normalizing them.

\subsection{The Fisher Graph Laplacian prior}\label{sec:prior}

A high-resolution source model can introduce many more degrees of freedom than the data can distinguish.
For example, a source with $N_s = 2048$ pixels on a side has millions of degrees of freedom, 
whereas the lensed image typically has a few thousand.
In that case, the Fisher matrix becomes low-rank and projection onto the nuisance tangent space becomes ill-defined.
To regularize the null space of the Fisher matrix, we introduce a prior.

For a source model represented on a Cartesian grid of pixels, a classical choice is gradient regularization, 
corresponding to a Gaussian prior with a nonzero covariance between neighboring pixels \citep[e.g.][]{Warren2003,Suyu2006}.
A convenient way to generalize this prior is to introduce a graph encoding correlations between pixels
and use the corresponding Laplacian matrix as the metric for a quadratic penalty \citep[e.g.][]{Pang2017}.
Here, we build on this idea by constructing the edges of the graph from the Fisher matrix
\begin{equation}
  \bm{s}^{\top} L \bm{s}
  =
  \frac{1}{2}
  \sum_{p\neq q}
  F_{pq} 
  \left(s_p-s_q\right)^2\, .
\end{equation}
The matrix $L$ represents a valid graph only when the off-diagonal elements of the Fisher matrix are non-negative.
In our context and specifically for the source model, 
this assumption is valid, such that the prior penalizes differences between source pixels in proportion to their mutual degeneracy.

A key property of this prior is that it diagonalizes the posterior metric 
\begin{equation}
  G = F + L = \mathrm{diag}(F \textbf{1})\, ,
\end{equation} 
where $\textbf{1}$ is the all-ones vector.
More generally, the strength of the prior is controlled by a Lagrange multiplier $\lambda$ 
corresponding to a smooth geometric transformation of the posterior metric.
The row-sums of the Fisher metric are an invariant of this transformation (see Appendix~\ref{sec:diagonalizing_fisher}),
such that the prior conserves a measure of the internal degeneracies of the source model.

\subsection{Matrix-free marginalization}\label{sec:matrix-free}

Computing the transfer function (defined in Section~\ref{sec:marginal})
requires applying the complement operator $P_{\eta}^{\perp}$ to each column of the substructure Jacobian.
For each substructure response $\bm{b}$, we compute its projection to the nuisance tangent space $P_{\eta}\bm{b}$ 
by solving a least-squares problem (see Section~\ref{sec:geometry}).
Using the projected response, we then form the complement as $P_{\eta}^{\perp}\bm{b} = \bm{b} - P_{\eta}\bm{b}$.

To solve the normal equation, we use the conjugate-gradient method \citep{Hestenes1952},
which only requires the action of the linear system on trial vectors.
Since matrix-vector products can be evaluated using automatic differentiation,
the action of the linear system can be evaluated with only a few forward passes of the simulator,
thus bypassing the need to store large matrices.
To ensure numerical stability of the solver, we use preconditioning.
Details about the matrix-free constructions of the various operators can be found in Appendix~\ref{sec:mf}.

When including all components of the nuisance model, we use an uninformative prior for the macro-model. 
Because the macro-model is low-dimensional, its contribution to the linear system can be evaluated using dense matrices. 

\section{Results}\label{sec:results}

Unless stated otherwise, all results use the fiducial \textit{arcs} configuration described in Section~\ref{sec:macro}. 
Lensed images are simulated using the adaptive ray-tracing scheme introduced in Section~\ref{sec:adaptive}, 
with a $3\times 3$ quadrature rule in each detector pixel. 
The fiducial source model uses a Cartesian grid with $N_s=2048$ pixels on a side, 
which we treat as an approximation to the continuum-source limit.
Throughout, we assume a diagonal data-space metric, $C^{-1} = \bbone$.
The reported sensitivity and signal-to-noise ratios (SNR) are thus relative quantities 
where the amplitude of the noise has been factored out.
In the context of this work, these quantities should be carefully interpreted as diagnostics, 
not as forecasts for a specific observing program.

\subsection{Simulator response to annulus eigenfunctions}

\begin{figure*}[thb!]
  \centering
  \includegraphics[width=0.75\textwidth]{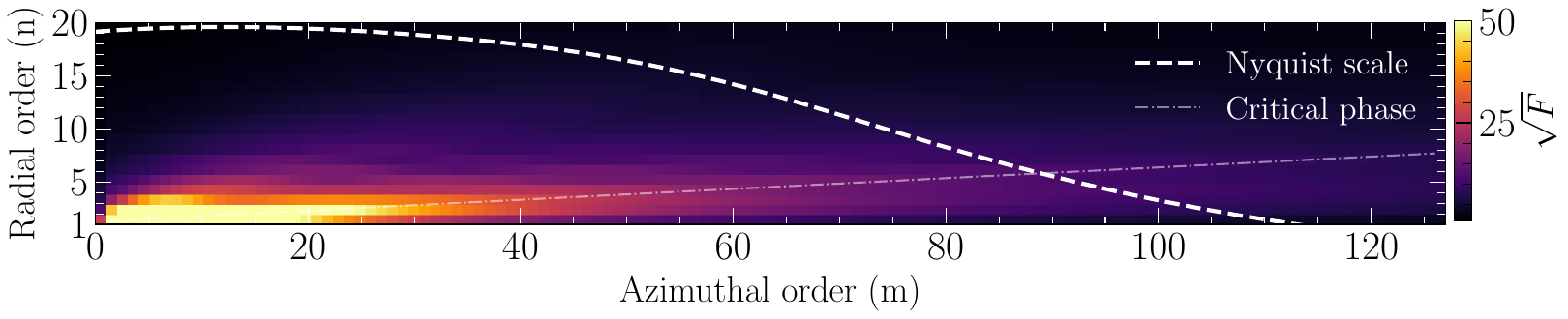}
  \caption{Simulator response to perturbations generated by the annulus eigenfunctions in the fiducial \textit{arcs} configuration.
    The Nyquist wavenumber curve and critical-phase relation are smoothed and plotted only as a visual aid.
  }
  \label{fig:fisher_information}
\end{figure*}

We first examine the response of the simulator to perturbations of the lensing potential. 
Figure~\ref{fig:fisher_information} shows the amplitude of the response associated with different annulus eigenfunctions, summed over the mode parity. 
We find that the response is most pronounced in an elongated band of modes that extends preferentially in the azimuthal direction, 
while dropping rapidly with radial order.

This band structure has a simple physical interpretation since near the tangential critical curve, 
the lens magnifies the source most strongly in the azimuthal direction. 
As a result, the simulator responds more sensitively to modes that generate azimuthal fluctuations near the Einstein radius, 
while modes with rapidly varying radial structure couple less efficiently to the lensed image.
This gives rise to an approximate critical phase relation supporting our interpretation (see Appendix~\ref{sec:wkb}).

\subsection{Degeneracies with the macro-model}\label{eq:macro_transfer}

Next, we aim to isolate the degeneracy between the macro-model and substructure.
To that end, we temporarily fix the source parameters to their ground truth values.
Since the macro-model only has 14 parameters,
marginalization reduces to a small least-squares problem which we solve using a standard Cholesky decomposition.

\begin{figure}[tb!]
  \centering
  \begin{tikzpicture}
    \node at (0, 0) {\includegraphics[width=0.45\textwidth]{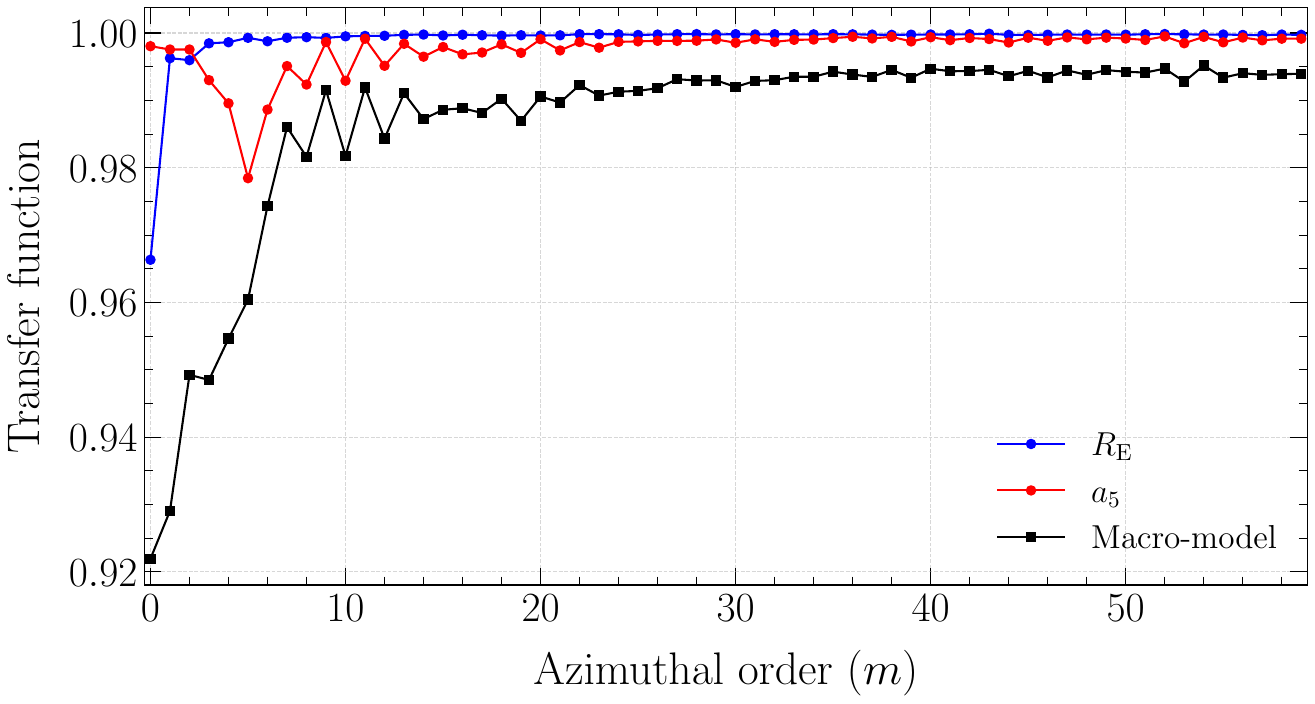}};
    \node[color=blue] at (-1.25, 0) {$J_{R_{\mathrm{E}}}$};
    \node at (-1.25, 0) {\includegraphics[width=2.5cm]{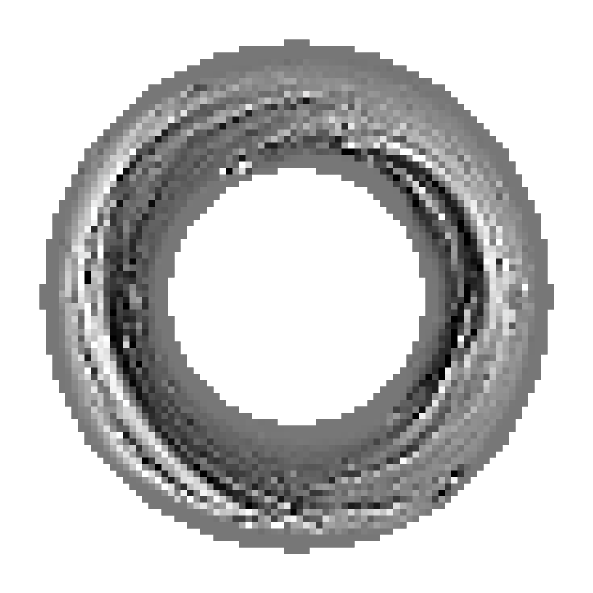}};
    \node[color=red] at (1, 0) {$J_{a_{5}}$};
    \node at (1, 0) {\includegraphics[width=2.5cm]{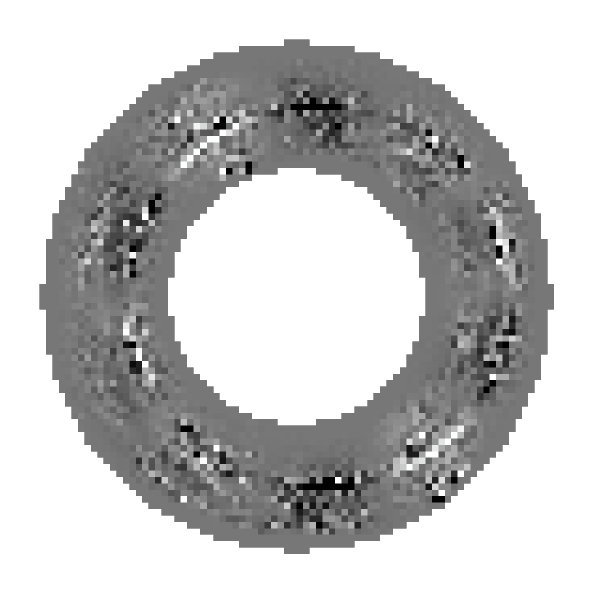}};
  \end{tikzpicture}
  \caption{Transfer function for the macro-model parameters in the \textit{arcs} configuration.
    The blue and red curves represent the transfer associated with individual parameter responses shown in the inset, 
    corresponding to the Einstein radius and the $\ell = 5$ multipole amplitude respectively.
    The black curve represent the joint transfer function associated with all 14 parameters.
    Only the first 60 azimuthal orders are shown to highlight the narrow band feature at small values of $m$.
  }
  \label{fig:absorbed}
\end{figure}

Figure \ref{fig:absorbed} shows the transfer function for the macro-model,
averaged over the radial orders and mode parity.
The most important absorption of anomalies occurs in a relatively narrow band of modes at low azimuthal order,
where the transfer function dips by a few percent below $T = 1$.
Outside this region, the transfer function rapidly reaches an asymptote close to $T=1$,
indicating that most eigenfunctions are not strongly degenerate with the macro-model.
By varying the lens configuration, we find that the amount of absorbed information 
does not significantly vary, 
indicating that this result is not strongly affected by the lens configurations considered here.

\subsection{Degeneracies with the source}

\begin{figure*}[t!]
  \centering
  \includegraphics[width=0.8\textwidth]{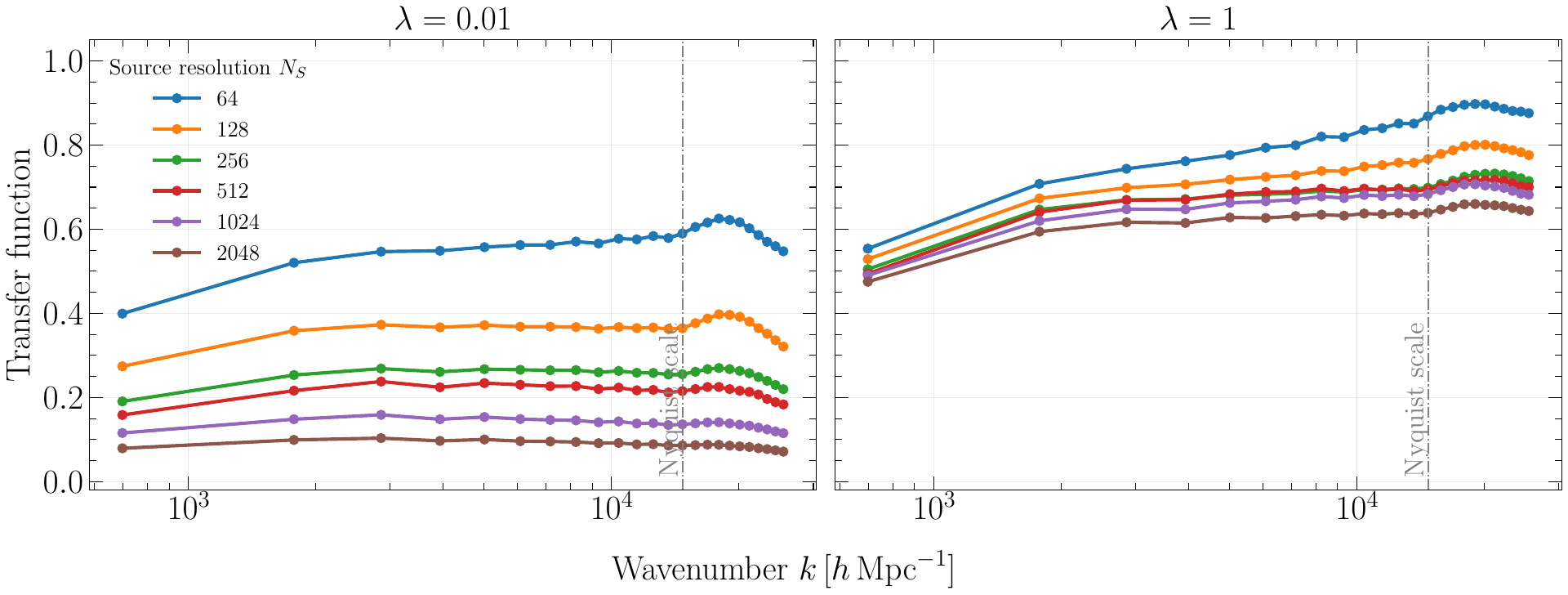}
  \caption{Transfer function corresponding to the marginalization of the full nuisance model in the \textit{arcs} configuration.
    The left panel corresponds to the weak-prior regime where the source is highly flexible, 
    while the right panel corresponds to the regularization scale at which the FGL prior diagonalizes the posterior metric.}
  \label{fig:resolution_sweep}
\end{figure*}

We now turn to degeneracies with the source model.
In this experiment, each substructure response is projected against the full nuisance tangent space, 
including both the macro-model and source degrees of freedom.
We explore the effect of increasing the dimensionality of the nuisance model
by repeating the experiment with different source resolutions.
The results are summarized in terms of the transfer function binned as a function of the wavenumber.
For convenience, the wavenumbers are converted to physical units using the angular diameter distance \citep[for a review, see][]{Hogg1999}.

In the left panel of Figure~\ref{fig:resolution_sweep}, we show the transfer function computed in the weak-prior regime.
In this regime, the regularization scale is sufficiently small for the source model to remain highly flexible,
while still allowing the marginalization algorithm to converge reliably.
We observe that, as the source resolution is increased, 
the amplitude of the transfer function decreases significantly across a broad range of wavenumbers.
For high-resolution models, the transfer function is strongly suppressed and approaches $T \simeq 0$, 
implying that inferred sensitivity to substructure is significantly affected by the choice of resolution for the source model.
Such a phenomenon aligns with previous empirical studies where it was found that varying the expressivity of the source model 
has a significant impact on the statistical significance of substructure detections \citep{Nightingale2024,Ephremidze2025}. 

The right panel of Figure~\ref{fig:resolution_sweep} shows the transfer function obtained when using the FGL prior at full strength. 
While a dependence on the source resolution remains, it is significantly dampened in this regime.
For the $N_s = 2048$ source model, the average transfer across wavenumbers stabilizes at $T \simeq 0.6$.
The contrast between the two panels demonstrates that the inferred sensitivity to substructure strongly depends 
on the regularization strength of the source model.
            
\subsection{Sensitivity to dark matter signals}

\begin{figure}[t!]
  \centering
  \begin{tikzpicture}
    \node at (0, 0) {\includegraphics[width=0.45\textwidth]{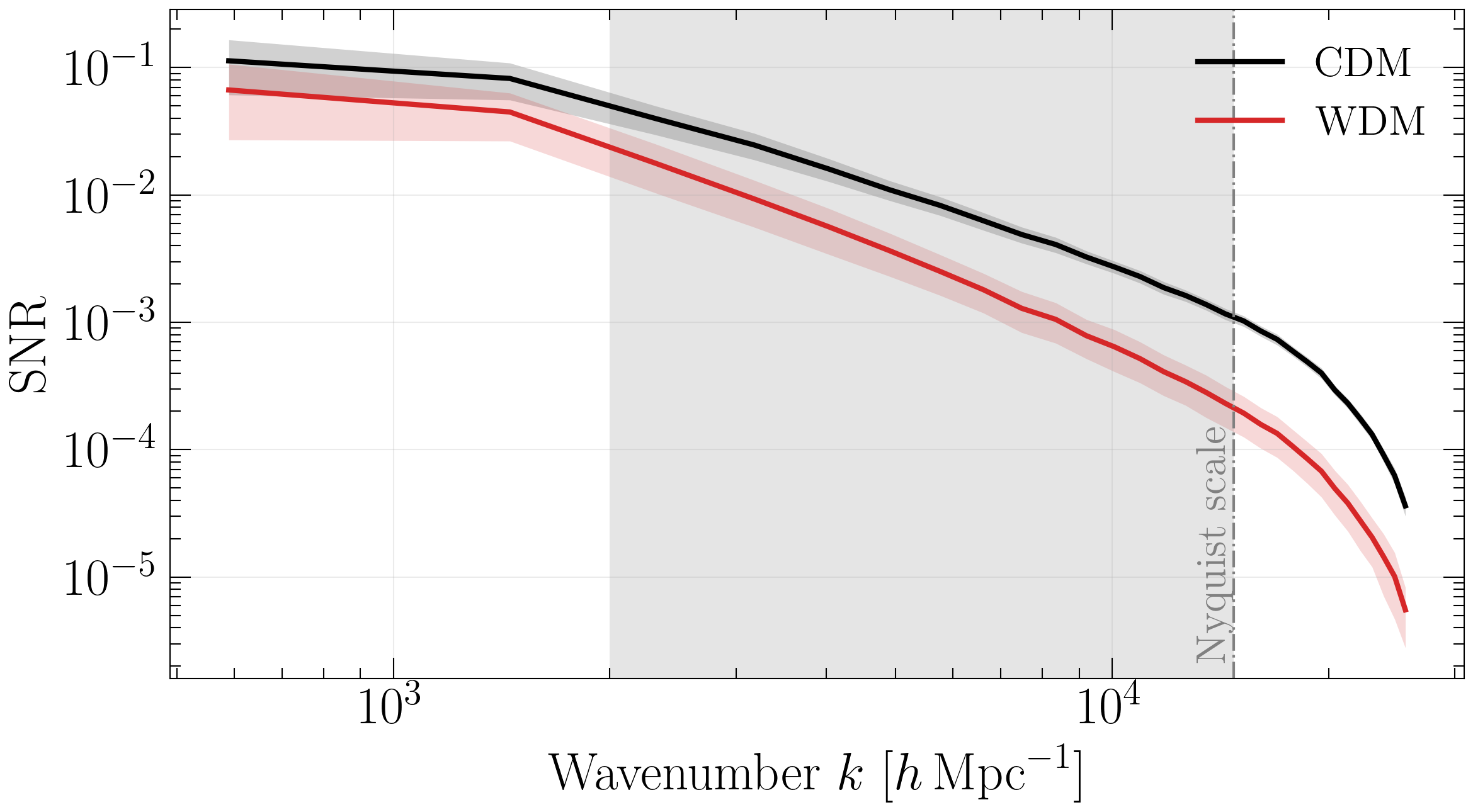}};
    \node at (-1.8, -0.3) {\includegraphics[width=2.5cm]{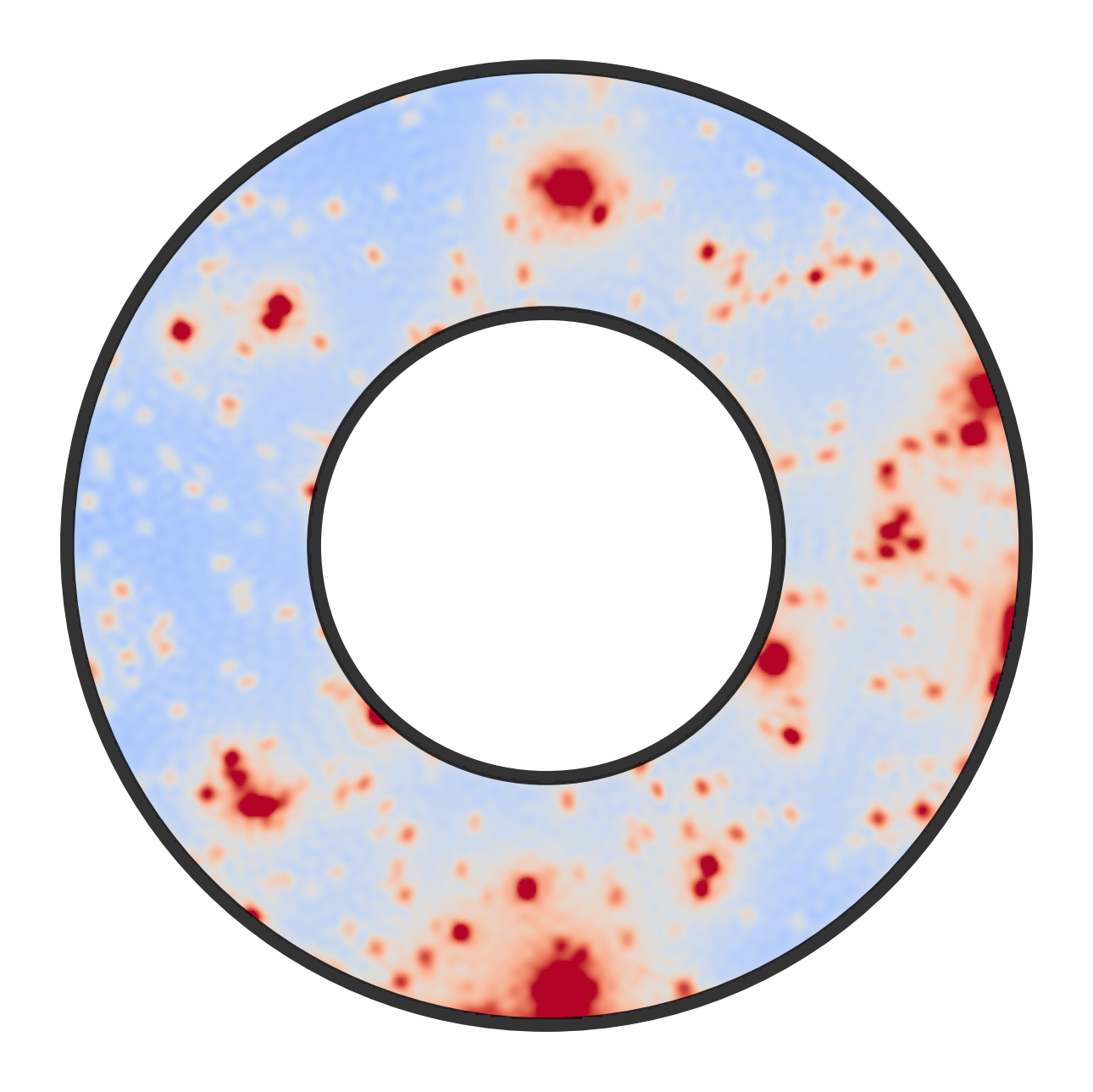}};
    \node at (0.45, -0.3) {\includegraphics[width=2.5cm]{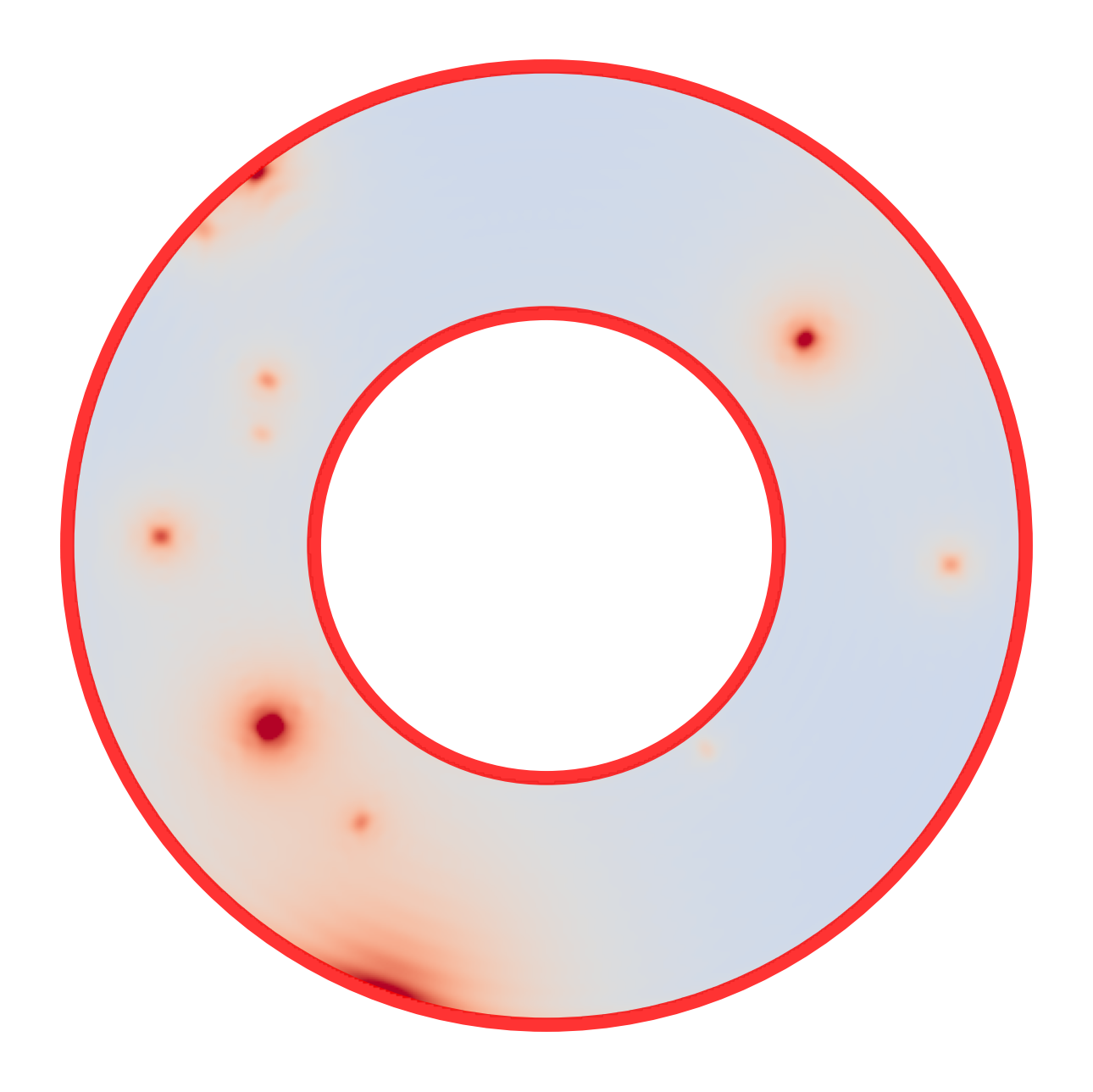}};
  \end{tikzpicture}
  \caption{SNR (in arbitrary units) of the coefficients generated by two different dark matter population models.
    The solid curves show the mean SNR for 100 realizations and the shaded contours represent the $1 \sigma$ scatter.
    Examples of substructure convergence projected in the annulus eigenbasis are shown in the inset 
    with a color scale indicating convergence contrast around $\kappa_{\mathrm{sub}} = 0$.
    The frequency window where the population signals are most distinguishable is highlighted in a gray shaded area.
  }
  \label{fig:snr}
\end{figure}

Having diagnosed the impact of the source resolution and the regularization strength on substructure sensitivity,
we now propagate one explicit configuration of the transfer function into dark matter signals.
Specifically, we choose $N_s = 2048$ and $\lambda = 1$ as our fiducial configuration.
Since the reported amplitude of the signals are conditional on this choice,
our observations will remain confined to the spectral content of the different population models.
To that end, we generate coefficients from the concordance CDM model and the alternative WDM model 
associated with a suppression of structure formation below subhalo mass $m_{\mathrm{hm}} = 9 \times 10^{7}\, M_{\odot}$ (see Appendix~\ref{sec:population_model}).
The inferred sensitivity to substructure mode $a$ after marginalizing the nuisance model is computed using the transfer function
\begin{equation}
  \mathcal{S}_a = T_a^{2}F_{aa} c_a^2\, .
\end{equation} 
We bin the sensitivity in wavenumber and define $\mathrm{SNR}(k)= \sqrt{\langle\mathcal{S}_a\rangle_k}$.

Figure~\ref{fig:snr} shows the SNR for the two population models.
At low-frequencies, the signals overlap because they are dominated by the high-mass end of the subhalo population, 
where both the CDM and WDM mass functions agree.
At higher frequencies, however,
they can be distinguished because of the suppression of low-mass subhalos in WDM realizations.
This difference is also evident in the power spectrum 
\begin{equation}
  \mathcal{P}_a = \langle c_a^{2} \rangle\, .
\end{equation} 
When fitting a power-law to the power spectrum, $\mathcal{P} \propto k^{-\alpha}$, 
in the frequency window highlighted in gray, we find $\alpha=2.7$ for CDM and $\alpha = 3.5$ for WDM.
The SNR for both population models is heavily suppressed at frequencies larger than the detector Nyquist scale, 
indicating that the simulator response vanishes for these modes.
The existence of a frequency window where the population signals could be distinguished is 
consistent with earlier works demonstrating that the matter power spectrum can be recovered from strong lensing data \citep{Hezaveh2016PS}.

\subsection{Detector sensitivity to substructure}

\begin{figure}[t!]
  \centering
  \begin{tikzpicture}
    \node at (0, 0) {\includegraphics[width=2.5cm]{arcs_lensed_image}};
    \node at (0, -2.5) {\includegraphics[width=2.5cm]{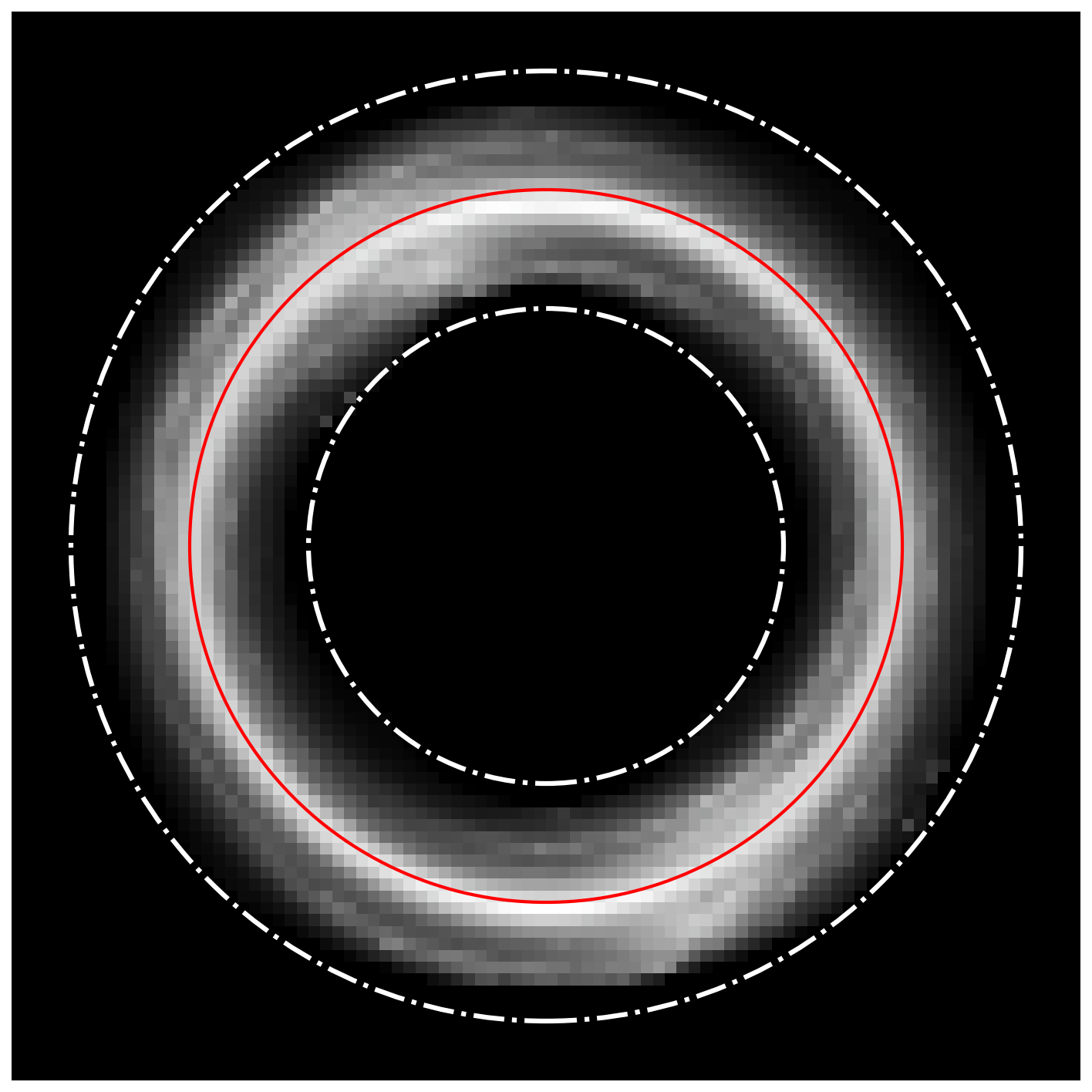}};
    \node at (0, -5) {\includegraphics[width=2.5cm]{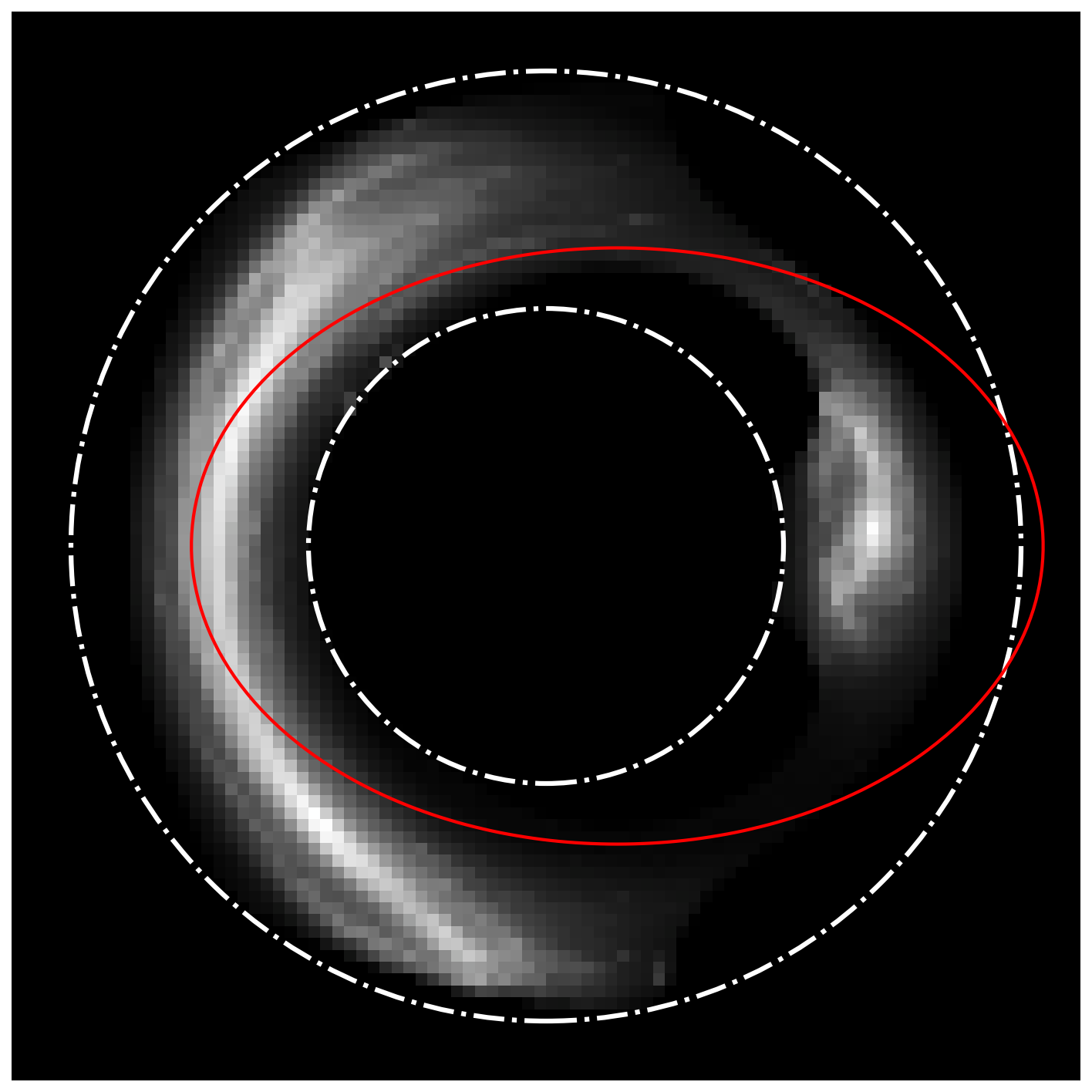}};
    \node at (0, -7.5) {\includegraphics[width=2.5cm]{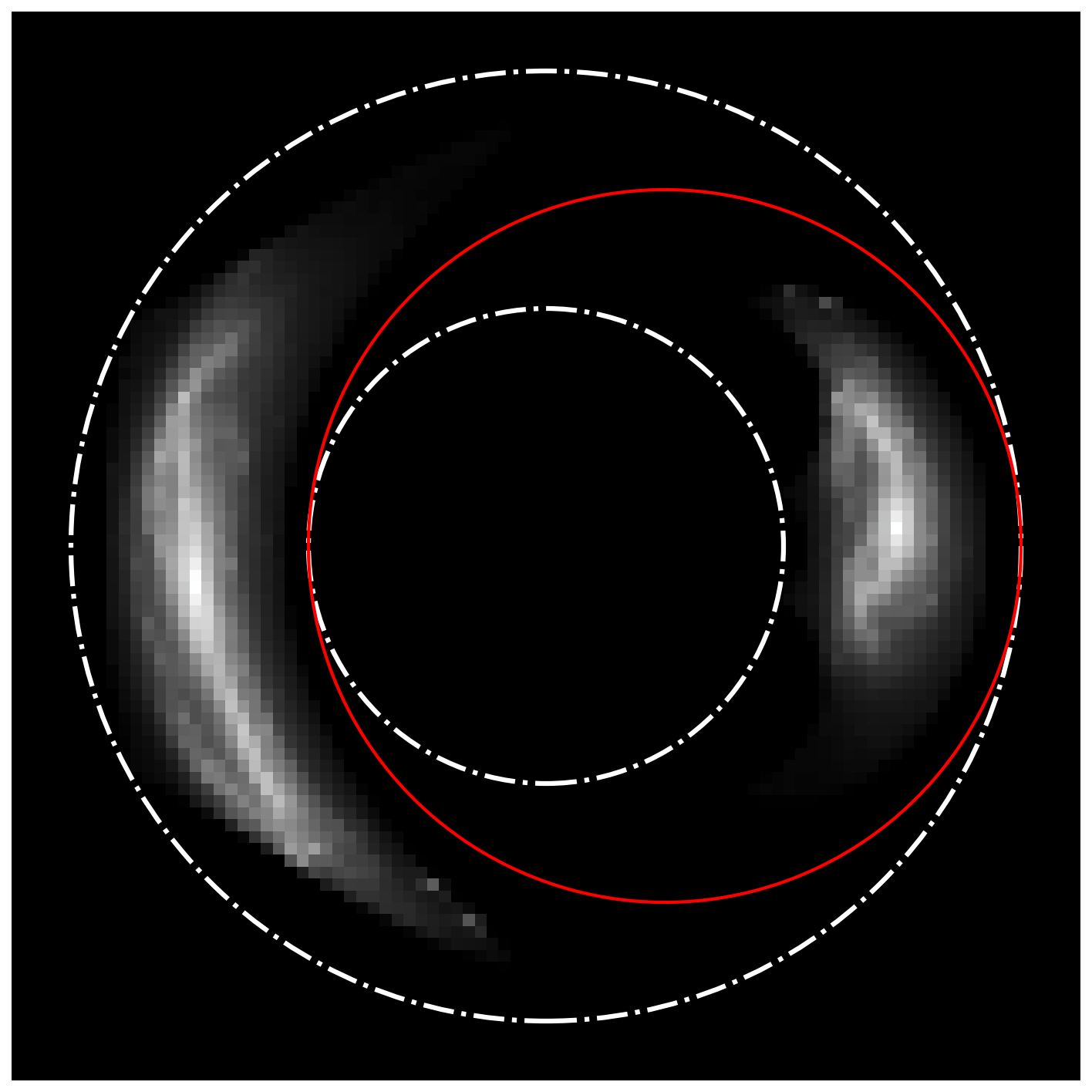}};
    \node at (2.5, 0) {\includegraphics[width=2.5cm]{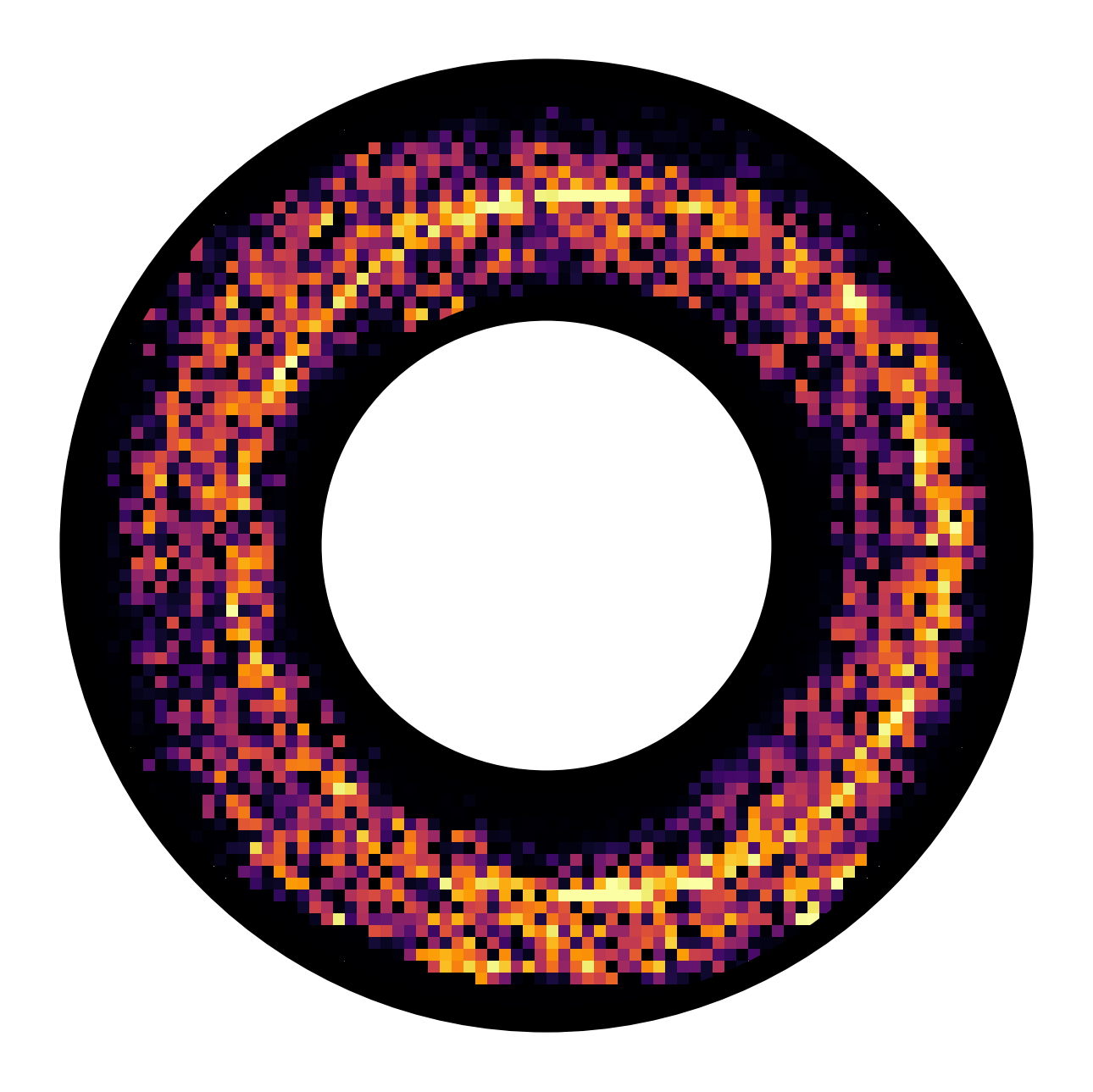}};
    \node at (2.5, -2.5) {\includegraphics[width=2.5cm]{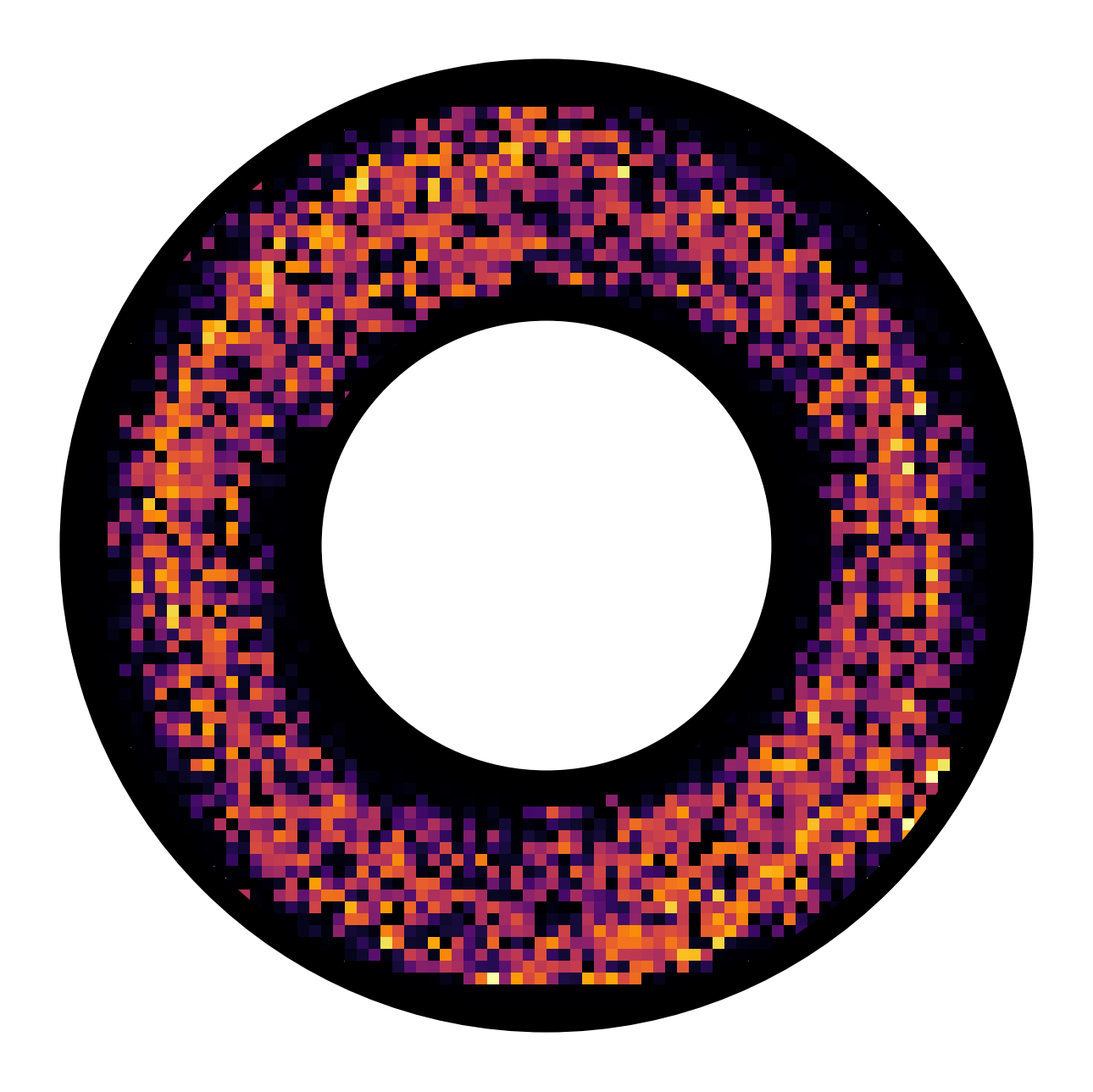}};
    \node at (2.5, -5) {\includegraphics[width=2.5cm]{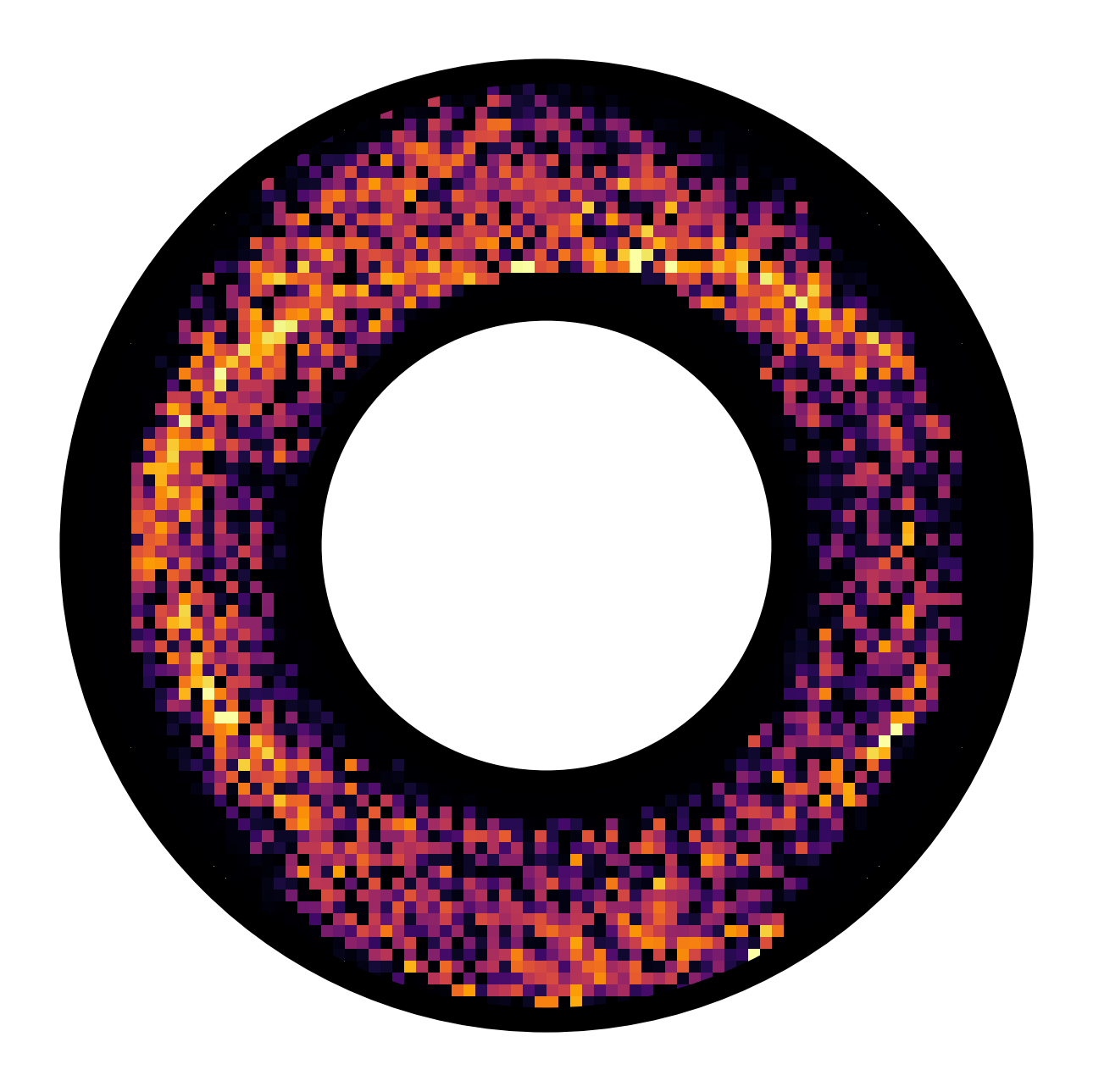}};
    \node at (2.5, -7.5) {\includegraphics[width=2.5cm]{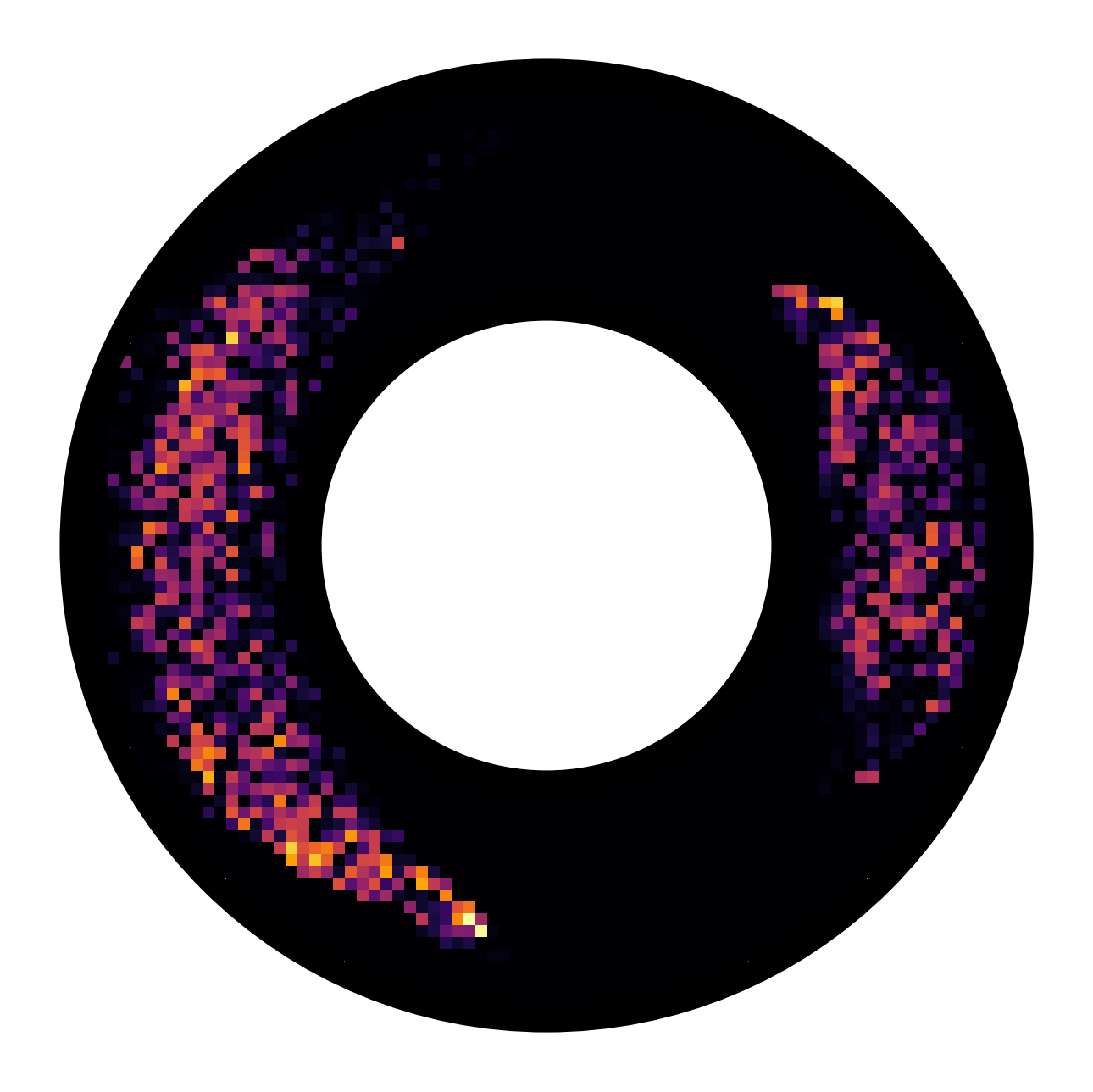}};
    \node at (5, 0) {\includegraphics[width=2.5cm]{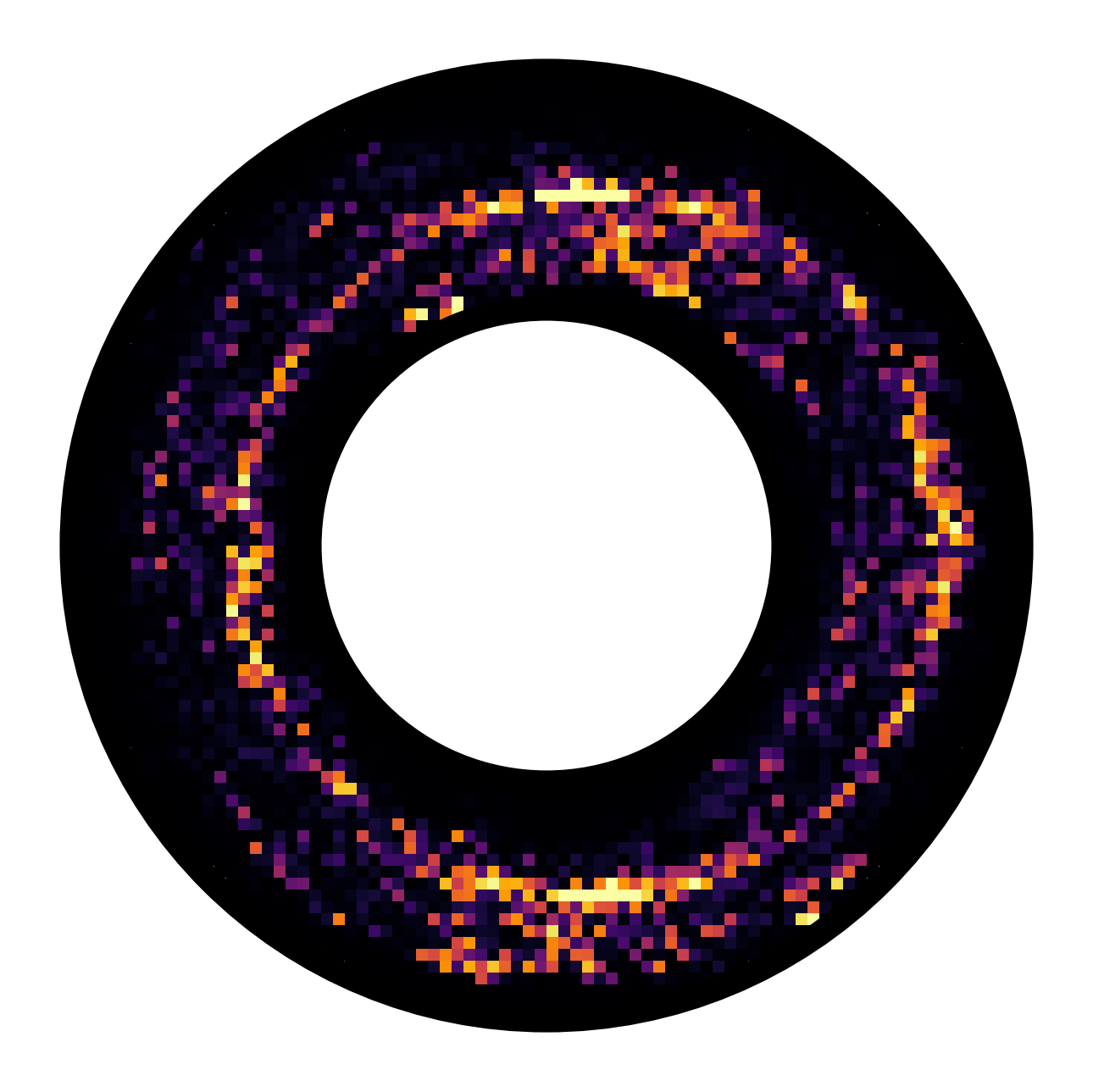}};
    \node at (5, -2.5) {\includegraphics[width=2.5cm]{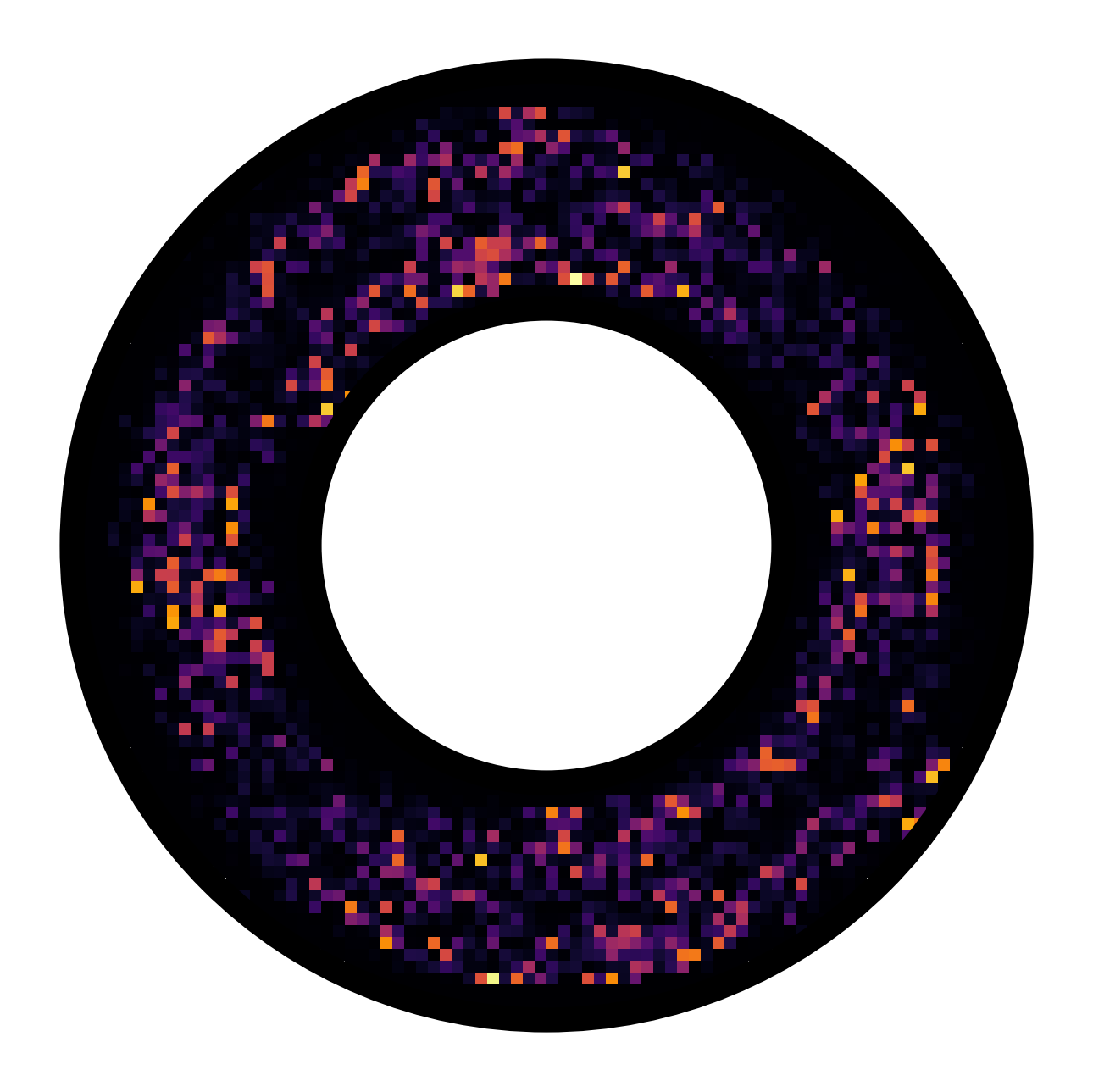}};
    \node at (5, -5) {\includegraphics[width=2.5cm]{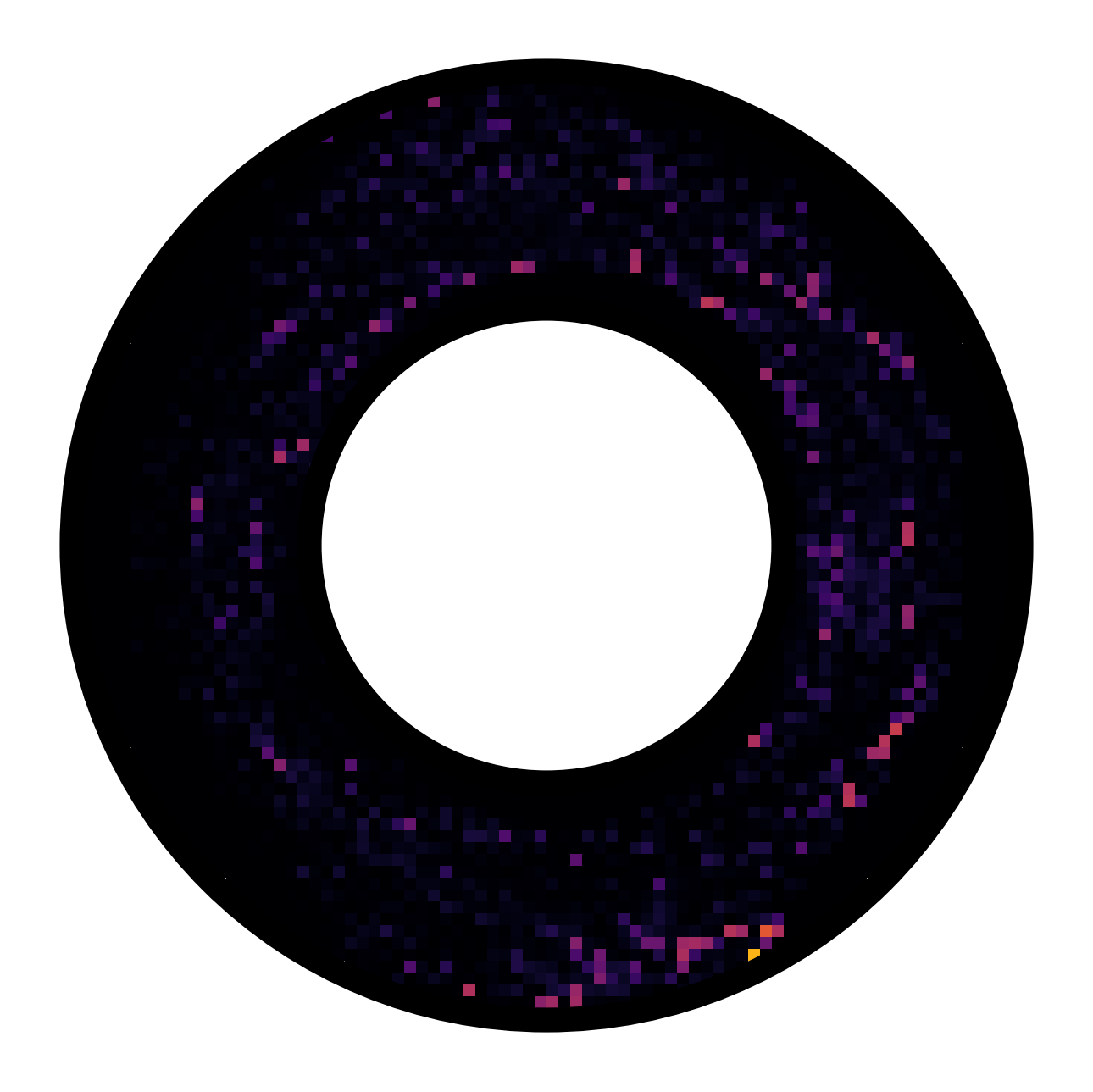}};
    \node at (5, -7.5) {\includegraphics[width=2.5cm]{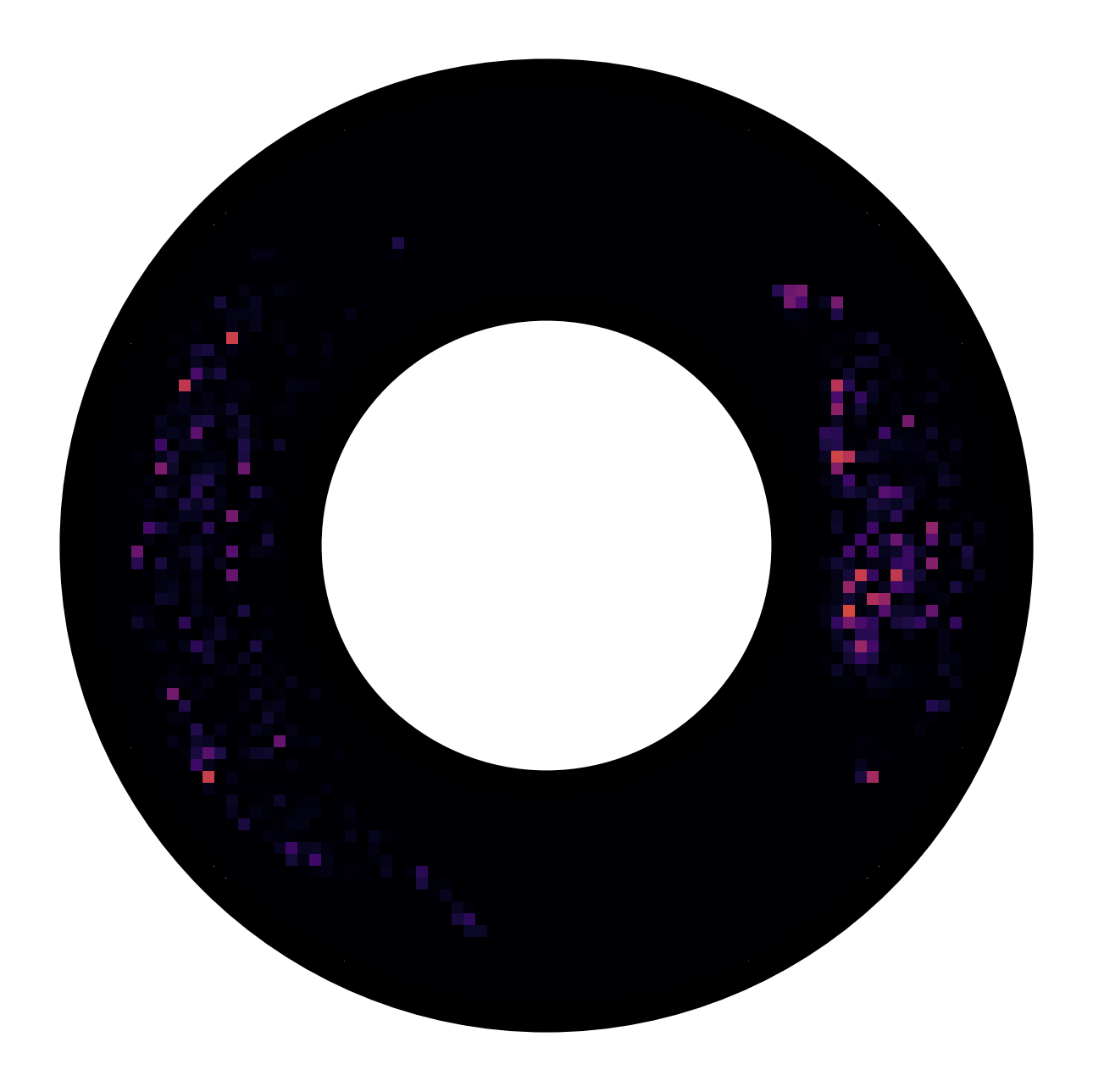}};
    \node[color=white] at (0, 0) {\textit{Arcs}};
    \node[color=white] at (0, -2.5) {\textit{Ring}};
    \node[color=white] at (0, -5) {\textit{Quad}};
    \node[color=white] at (0, -7.5) {\textit{Double}};
    \node at (2.5, -9) {Before};
    \node at (5, -9) {After};
  \end{tikzpicture}
  \caption{
    The detector response before marginalization for a CDM population of subhalos is shown in the middle column. 
    The rightmost column corresponds to the detector sensitivity after marginalization of nuisance parameters.
    The corresponding lensed-image is shown in the leftmost column, where the critical curve is plotted as a red curve for all lens configuration.}
  \label{fig:pixel_sensitivity}
\end{figure}

As a byproduct of previous experiments, we now inspect the detector sensitivity to substructure.
We define the sensitivity of a detector pixel $i$ as the marginalized response to the coefficient power spectrum
\begin{equation}
  \mathcal{R}_i = \sum_a \mathcal{P}_{a} (\tilde{J}^{\perp}_{ia})^{2}\, ,
\end{equation} 
where $\tilde{J}^{\perp}_a = C^{-1/2}P^{\perp}_{\eta}J_a$ is a shorthand to denote a whitened projected column of the substructure Jacobian.

Figure~\ref{fig:pixel_sensitivity} shows the detector sensitivity before and after marginalization for all lens configurations. 
The \textit{arcs} configuration is by far the most sensitive along a circle mapping the critical curve, 
showing that such pixels are more robust to marginalization.
The pixels near the critical curve in the \textit{ring} configuration are not particularly sensitive, 
either before or after marginalization.
For this particular configuration, this can be explained by observing that pixels near the critical curve
map to the core of the source galaxy where its surface-brightness is locally smooth.
The \textit{double} configuration is the least sensitive because the lensed image does not overlap with the critical curve.
In the case of the \textit{quad} configuration, we find that despite a large response near the critical 
curve before marginalization, the sensitivity after marginalization drops significantly for pixels near the critical curve.

\section{Discussion}\label{sec:discussion}

\subsection{On the macro-substructure degeneracy}

It has long been known that the macro-substructure degeneracy is a significant confounding factor for 
determining the true origin of flux anomalies in strongly lensed quasar data \citep[e.g.][]{Mao1998,Dalal2002,Hsueh2016,Gilman2017}.
For highly-magnified arcs, the degeneracy is less well understood, 
though recent empirical studies have suggested that the angular complexity of the macro-model
can still be degenerate with parameters describing substructure, 
specifically for substructure parameters describing the properties of an individual perturber \citep{ORiordan2024}.

Our analysis offers a complementary view by studying this degeneracy in the spectral domain, 
where substructure is represented by perturbations of the lensing potential.
Since we use radial boundary conditions on the annulus, the spectrum of the perturbations is truncated,
such that we do not account for modes associated with scales larger than the annulus.
For such low-frequency modes, previous studies found that degeneracies with the macro-model could be more significant \citep{CyrRacine2019}. 
Here, we find that macro-model parameters are not significantly degenerate with high-frequency perturbations.
This result is largely expected since the macro-model describes a smooth potential function.
While multipoles in the macro-model can introduce some degeneracies with modes corresponding to its azimuthal harmonics,
this degeneracy is largely confined to eigenfunctions with small radial orders, such that overall degeneracy with substructure remains small.

\subsection{Source expressivity as an information sink}

In the high-frequency substructure regime, the degeneracy with the source model is a much greater concern.
In particular, source models with a coarse resolution (e.g.~$N_s = 64$) are found to have large transfer compared to high-dimensional models (e.g.~$N_s = 2048$).
This result shows that the source resolution is not a neutral modeling choice. 
By restricting the set of source configurations available to the fit, 
a coarse grid can artificially increase the apparent sensitivity to dark matter perturbations.

The opposite failure mode appears when the source is made highly expressive and weakly regularized.
As the source resolution is increased, the transfer becomes small, 
indicating that substructure perturbations are strongly degenerate with the source parameters.
This observation is consistent with empirical studies showing that
high-dimensional source models could explain nearly all of the data in a sample of 30 strong lenses, even when excluding substructure in the analysis \citep{Legin2025}.
Finding a physically motivated regularization mechanism for the source 
is therefore a central bottleneck in extracting signals from an unresolved dark matter population.

\subsection{Regularization near caustics}

\begin{figure*}[t!]
  \centering
  \includegraphics[width=\textwidth]{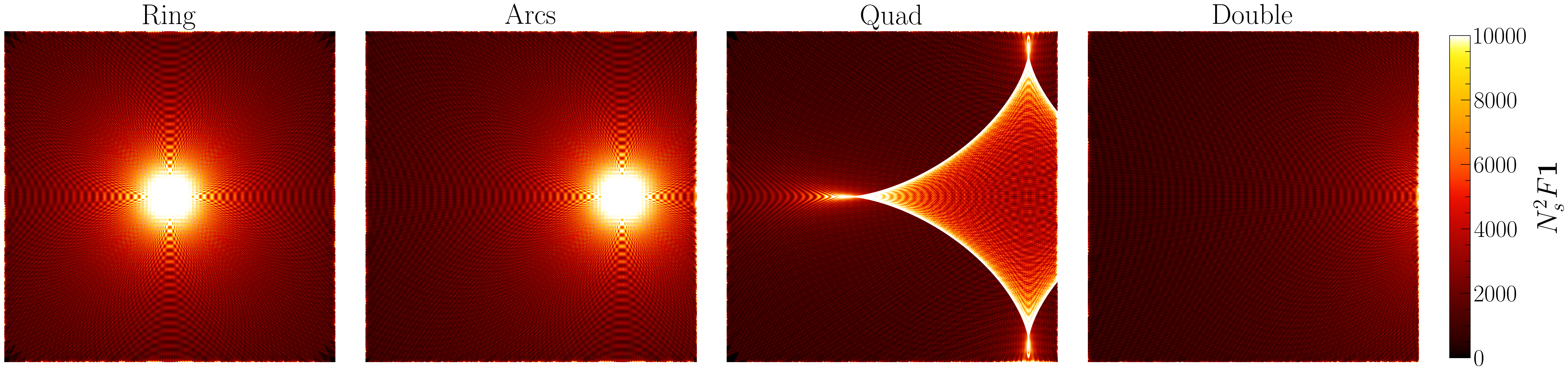}
  \caption{
  Degree of the Fisher matrix for the $N_s = 2048$ source model for all lens configurations.
  For interpretability, the color scale is saturated at large values.
}
  \label{fig:degree}
\end{figure*}

In this work, we introduced the FGL prior as a diagnostic tool to study how source regularization can affect inferred dark matter sensitivity. 
To interpret the action of this prior, it is useful to inspect how the Fisher matrix distributes information across the source plane. 
A simple diagnostic is the degree of the Fisher matrix
\begin{equation}
  (F \textbf{1})_{p} = F_{pp} + \sum_{q \neq p} F_{pq}\, ,
\end{equation} 
This quantity contains two contributions.
The diagonal term, $F_{pp}$, measures the direct sensitivity of the data to source pixel $p$, 
while the off-diagonal sum measures how strongly it is degenerate to other source pixels. 

In Figure~\ref{fig:degree}, we plot this quantity for all lens configurations,
where it can be observed that pixels with a large degree are concentrated near caustics.
This is expected as these source-plane regions are associated with high magnification, such that small-scale gradients in the surface-brightness 
of the source can contribute to extended regions of the lensed image.
By inspecting the contribution from the diagonal component separately, we find that its spatial structure is correlated with the magnification map, as expected.
However, the off-diagonal contribution is much larger than the diagonal,
especially for high-dimensional models (e.g.~$N_s = 2048$),
and has a spatial structure that is also strongly correlated with the magnification map.
This indicates that source pixels near the caustics have degenerate responses in the data.

By targeting these degenerate pixels and penalizing relative variations between them, the FGL prior is able to reduce 
the effective expressivity of the source model in those regions, 
thus reducing its ability to absorb anomalies during marginalization.
This interpretation is consistent with the transfer function results.
When the strength of the FGL prior is relaxed ($\lambda = 0.01$), 
the source model gains access to more degrees of freedom and can absorb a larger fraction of the substructure response.
Conversely, when the prior is applied at full strength ($\lambda = 1$), 
part of this absorption channel is suppressed, and more of the substructure response remains after marginalization.

We emphasize, however, that these results do not imply that the FGL prior is the unique way to restore sensitivity to substructure.
Other priors, even as simple as a scalar Tikhonov regularization,
would also reduce the effective expressivity of the source model and could produce qualitatively similar results.
The value of the FGL construction is instead as a diagnostic,
showing that regulating the internal degeneracies of the source model is sufficient for restoring substructure sensitivity.

To gain more insights into this mechanism, we can also inspect the sensitivity maps of the detector.
For the \textit{arcs} configuration, the sensitivity is largest near the critical curve,
where the degree map indicates that the corresponding source pixels have very similar responses in the data.
For the \textit{quad} configuration, however, the opposite is true since marginalization strongly suppresses the sensitivity near the critical curve.
This may indicate that the effect of FGL regularization depends on the caustic geometry and on how the macro-model shapes the source-plane information geometry.
A more detailed study of this dependence is left for future work.

Overall, these results provide a complementary perspective on adaptive source reconstruction methods \citep{Vegetti2009,Nightingale2015}.
Adaptive methods use the lens mapping to allocate more degrees of freedom in regions of high-magnification, 
thus increasing the expressivity of the model in proportion to the local density of rays.
Since our analysis shows that high-magnification regions is also where the source model can develop significant internal degeneracies,
this suggests that regularization of degrees of freedom near caustics, 
as occurs in adaptive methods that impose gradient priors, 
can play an important role in controlling the amount of substructure information that survives marginalization.

\section{Conclusion}\label{sec:conclusion}

In light of our results, we conclude that detecting the collective signal of an unresolved dark matter population 
in strong lensing data will require precise statistical control over the nuisance models used in the inference.
Untangling substructure from expressive nuisance models inevitably exposes degeneracies between them, 
such that even seemingly minute modeling choices can have significant consequences for the inferred statistical significance of dark matter signals
\citep[e.g.][]{Ritondale2019,Despali2022,ORiordan2024,Nightingale2024,Ephremidze2025,Filipp2025}.

In this work, we developed a framework to diagnose these effects, 
where we represented substructure perturbations to the lensing potential in a spectral basis defined on an annulus surrounding the lensed image.
We then used a differentiable strong-lensing simulator to map the degeneracies of these perturbations with expressive nuisance models,
including a macro-model with high-order multipoles and a high-dimensional model for the background source.
Using tools from information geometry, we characterized the transfer function associated with the marginalization 
of the nuisance models while varying the expressivity of the source model by increasing its resolution and tuning the regularization strength.

Our results show that degeneracies with the macro-model are relatively small and confined to low-frequency substructure modes,
while degeneracies with the source model are much more broad in scale and can drastically impact the inferred sensitivity of dark matter signals.
This effect becomes stronger as the expressivity of the source model is increased, 
demonstrating that source resolution and regularization are not neutral modeling choices.

To study the regularization of the source, we introduced the Fisher Graph Laplacian prior as a diagnostic tool that
uses the internal degeneracies of the source model as a leverage over the transfer function for substructure.
With this construction, we showed that strong degeneracies exists between pixels near regions of high-magnification in the source plane, 
thus highlighting how important the regularization of those degrees of freedom is.

As a demonstration of how the framework can be used, we then propagated one fiducial transfer function into substructure 
realizations generated from CDM and WDM subhalo populations.
Although conditional on this particular choice, the calculation showed that the population models make distinct predictions 
in a finite frequency window where the WDM suppression of low-mass subhalos is associated with a lower SNR and substructure coefficient power.

The cornerstone of this framework is a fully differentiable simulator,
offering a transparent architecture where the sensitivity to underlying model parameters can be tracked via exact gradients.
As a byproduct of the recent deep learning paradigm shift in computer science \citep{LeCun2015},
it is now possible to build pipelines on top of existing computer languages that natively support automatic differentiation, 
such as \texttt{PyTorch} \citep{torch} and \texttt{JAX} \citep{jax}.
Since an emerging generation of open-source strong-lensing software packages has begun utilizing this infrastructure \citep{Gu2022,Stone2024,Huang2026},
a window of opportunity is now open for exploiting these tools,
the potential of which can be gauged by looking at other areas of computational astrophysics that have started taking advantage of differentiable simulators
\citep[e.g.][]{Pope2021,Zeghal2022,Li2022,Champagne2023,Desdoigts2024,Desdoigts2025,Charles2026}.

\section{Acknowledgements}
I would like to thank Laurence Perreault-Levasseur for her invaluable support and guidance.
Without her encouragements, it is likely I would not have finished this work.
I am also grateful for valuable discussions with Yashar Hezaveh and Neal Dalal.
This work was funded by an NSERC CGS D scholarship and is in part supported 
by computational resources provided by Calcul Quebec and the Digital Research Alliance of Canada.

Generative AI tools, namely ChatGPT and Codex (OpenAI), were used during the development of this work as interactive research and programming aids. 
Their use included code generation, exploration of mathematical derivations, discussion of scientific interpretations, 
and occasional assistance with the organization of the manuscript. 
The research questions, methodological choices, numerical experiments, and final manuscript were developed and validated by the author. 
All mathematical or numerical results, references, and scientific claims included in the manuscript were reviewed and validated by the author, 
who takes full responsibility for the work.

Software used: \texttt{astropy} \citep{astropy:2013,astropy:2018}, \texttt{jupyter} \citep{jupyter}, \texttt{matplotlib} \citep{matplotlib}, 
\texttt{numpy} \citep{numpy}, \texttt{SciPy} \citep{scipy}, \texttt{PyTorch} \citep{pytorch}, 
\texttt{tqdm} \citep{tqdm}, \texttt{pandas} \citep{pandas}, \texttt{Caustics} \citep{Stone2024}

\bibliography{bib}
\appendix

\section{Annulus eigenfunctions with Neumann boundary conditions}\label{sec:neumann_bc}

The symmetry of the annulus domain leads to the factorisation of its eigenfunctions
\begin{equation}
\phi_{mnp}(r, \varphi) = R_{mn}(r) \Theta_{mp}(\varphi)\, .
\end{equation} 
The structure of each mode is labeled by a compound index $a=(m,n,p)$, including the azimuthal order $m$, 
the radial order $n$ and a parity $p \in \{\cos, \sin\}$ distinguishing the two Fourier modes for the angular component.

For each azimuthal order $m$, the radial component is fixed by the solution to an eigenvalue problem,
which takes the form of Bessel's equation.
Its general solutions are a linear combination of Bessel functions of the first and second kind
\begin{equation}
  R_{mn}(r) = A_{mn} J_m(k_{mn} r) + B_{mn}Y_{m}(k_{mn}r)\, .
\end{equation}
There are three quantities to determine: the relative amplitudes $A_{mn}$ and $B_{mn}$, and the wavenumber $k_{mn}$.
Imposing Neumann boundary conditions at the inner and outer radius yields 
\begin{align}
  \label{eq:det_1}
  A_{mn}J'_m(k_{mn}r_{\mathrm{in}}) + B_{mn}Y'_m(k_{mn} r_{\mathrm{in}}) &=  0\, ,\\
  \label{eq:det_2}
  A_{mn}J'_m(k_{mn}r_{\mathrm{out}}) + B_{mn} Y'_m(k_{mn} r_{\mathrm{out}}) &= 0\, ,
\end{align}
where $J'_m$ and $Y'_m$ are derivatives of Bessel functions of the first and second kind, respectively.
These two constraints, together with the normalization constraint
\begin{equation}
  \int_{r_{\mathrm{in}}}^{r_{\mathrm{out}}} R^{2}_{mn}(r) r \dd r = 
  \begin{cases}
    \displaystyle
    \frac{1}{2\pi}\, , & m = 0\, ; \\[1em]
    \displaystyle
    \frac{1}{\pi}\, , & \mathrm{otherwise}\, ,
  \end{cases}
\end{equation} 
fully specify the unknowns.
Since we can compute the normalization post facto, the relative amplitudes only need to account for the boundary condition.
We choose
\begin{equation}
  A_{mn} = 1 \, , \qquad B_{mn} = - \frac{J'_m(k_{mn} r_{\mathrm{in}})}{Y'_m(k_{mn} r_{\mathrm{in}})}\, .
\end{equation}

\section{Roots of the determinant equation}\label{sec:roots}

Non-trivial solutions for Neumann boundary conditions exist only when the determinant of these constraints vanishes.
This determinant can be written as a function of the wavenumber
\begin{equation}\label{eq:determinant}
  D_{m}(k) = J'_m(k r_{\mathrm{in}}) Y'_m(k r_{\mathrm{out}}) - J'_m(k r_{\mathrm{out}}) Y'_m(k r_{\mathrm{in}}) \, .
\end{equation}
For each azimuthal order $m$, there exists a discrete spectrum of wavenumbers $k_{mn}$, with $n \in \{1,\dots, n_{\rm max}\}$, 
which are roots of the transcendental equation $D_{m}(k) = 0$.
To find the first $n_{\rm max}$ roots of Equation~\eqref{eq:determinant}, we use a standard bracketing root-finding algorithm,
initialized with $Cn_{\rm max}$ equally-spaced points within the interval $[k_{\rm min}, k_{\rm max}]$.
$C = 10$ is a parameter chosen conservatively to find all roots within the interval.
We then look for brackets $(k_i, k_j)$ for which the sign of the determinant flips, i.e.~$D_m(k_i)D_m(k_j) < 0$.
Such pairs serve as initial conditions for Brent's method, implemented in \texttt{SciPy} \citep{scipy}, 
which is used to solve $D_{m}(k) = 0$ within the bracket, up to machine precision.

The lower bound for the search interval is set by the turning point
\begin{equation}
  k_{\rm min} = 
  \begin{cases}
    \displaystyle
    \frac{\sqrt{m^{2} - \frac{1}{4}}}{r_{\rm out}},& m > 0\, ;\\
    0, & m = 0\, ,
  \end{cases}
\end{equation} 
where the local wavenumber of Bessel's equation becomes real within the annulus.
For the upper bound, we impose a conservative quantization condition to the WKB phase integral
\begin{equation}\label{eq:quantization}
  S_{\rm WKB}(k_{\rm max}) = (n_{\rm max} + 1) \pi\, .
\end{equation} 
Details about this approximation are discussed in Appendix~\ref{sec:wkb}.
Equation~\eqref{eq:quantization} is solved once per azimuthal order $m$ using Brent's method.
The search for $k_{\rm max}$ is initialized with Equation~\eqref{eq:k_mn_wkb}.

\section{WKB approximation of Bessel functions}\label{sec:wkb}

The radial part of an annulus eigenfunction satisfies Bessel's equation,
\begin{equation}
  R''_{m} + \frac{1}{r}R'_{m} + \left( k^2 - \frac{m^2}{r^2} \right) R_{m} = 0\, .
\end{equation}
After a standard transformation, $R_m(r) = r^{-1/2}u(r)$, it takes the normal form of a wave equation 
\begin{equation}
  u'' + q^{2}(r) u = 0\,, \qquad \,\,\, q(r) = \sqrt{k^2 - \frac{m^2 - \frac{1}{4}}{r^2}}\, .
\end{equation} 
The function $q(r)$ is the local radial wavenumber of the mode. 
In the oscillatory region, the WKB phase accumulated across the annulus is
\begin{equation}
  S_{\rm WKB}(k) = \int_{r_{\rm in}}^{r_{\rm out}} q(r) \dd r\, .
\end{equation}
This phase is a monotonically increasing function of $k$ and provides an approximate way of counting radial oscillations.

It is useful to replace the slowly varying radial wavenumber by its value near a representative radius $r_\ast$. 
Locally, the mode then behaves like a wave whose total wavenumber is decomposed into angular and radial contributions $k^2 \simeq \left(\frac{m}{r_\ast}\right)^2 + k_r^2$.
Estimating the radial wavenumber by $k_r \sim \frac{n\pi}{r_{\rm out}-r_{\rm in}}$,
gives the approximate relation
\begin{equation}\label{eq:k_mn_wkb}
  k_{mn} \approx \left[ \left( \frac{m}{r_{\ast}}\right)^{2} + \left( \frac{ n \pi}{r_{\rm out} - r_{\rm in}} \right)^{2} \right]^{1/2}.
\end{equation}
This expression can be used as a back-of-the-envelope estimate of the wavenumber of an eigenfunction.

It can also be used to interpret the band structure of the simulator response, shown in Figure~\ref{fig:fisher_information}.
A mode whose wavenumber satisfies
\begin{equation}\label{eq:critical_phase}
  k_{mn} \approx \frac{m}{r_\ast}
\end{equation}
has little radial phase accumulation near that radius in the local WKB sense. 
When $r_\ast$ is chosen to be the Einstein radius, 
Equation~\eqref{eq:critical_phase} corresponds to modes whose oscillations are aligned with the tangential critical curve 
of a lens that has a circularly symmetric profile.
In the more general case, this estimate may not be as useful since the detailed shape of the substructure response ultimately depend on a host of factors, 
including the source morphology, the macro-model and the detector.

\section{Moment-matched kernels for adaptive ray tracing}
\label{sec:adaptive_kernel}

In Section~\ref{sec:adaptive}, we approximate finite collecting areas by a Gaussian kernels. 
This appendix gives the elementary derivation of the covariance factors used in that approximation.

\subsection{Subpixel collecting area}

Consider first a square image-plane subpixel of width $\Delta\theta$, centered
at the origin. If the subpixel is approximated as a uniform top-hat collecting area, its one-dimensional marginal distribution is
\begin{equation}
  p(\theta)
  =
  \frac{1}{\Delta\theta},
  \qquad
  -\frac{\Delta\theta}{2}
  \leq
  \theta
  \leq
  \frac{\Delta\theta}{2}.
\end{equation}
The variance of this distribution is
\begin{equation}
  \mathrm{Var}(\theta)
  =
  \frac{1}{\Delta\theta}
  \int_{-\Delta\theta/2}^{\Delta\theta/2}
  \theta^2\,\dd\theta
  =
  \frac{\Delta\theta^2}{12}.
\end{equation}
Since the square top-hat factorizes along the two image-plane coordinate
directions, its moment-matched covariance is
\begin{equation}
  \Sigma_{\theta}
  =
  \frac{\Delta\theta^2}{12}\,\bbone.
\end{equation}

Under the local linearized lens mapping around the ray position
$\bm{\theta}_q$, an image-plane displacement $\delta\bm{\theta}$ maps to a
source-plane displacement
\begin{equation}
  \delta\bm{\beta}
  =
  A \delta\bm{\theta},
\end{equation}
where $A = A(\bm{\theta}_q)$ is the lensing matrix.
Therefore, the covariance $\Sigma_{\beta}(\bm{\theta}_q)$ of the collecting area transported to the source-plane is given by
\begin{equation}
  \Sigma_{\beta}(\bm{\theta}_q)
  =
  A\Sigma_{\theta}A^\top\, .
\end{equation}

\subsection{Effective width of bilinear interpolation}

The source grid introduces an additional numerical smoothing scale commonly defined by the bilinear interpolation kernel.
To assign a Gaussian covariance to this operation, 
we match the second moment of an effective one-dimensional kernel.
In one dimension, linear interpolation corresponds to the triangular kernel
\begin{equation}
  T(x)
  =
  \begin{cases}
    1 - |x|/\Delta\beta, & |x| < \Delta\beta,\\
    0, & |x| \geq \Delta\beta,
  \end{cases}
\end{equation}
where $\Delta\beta$ is the spacing between source-plane pixels. Normalizing this kernel gives
\begin{equation}
  p(x)
  =
  \frac{1}{\Delta\beta}T(x).
\end{equation}
Its variance is
\begin{align}
  \mathrm{Var}(x)
  &=
  \frac{1}{\Delta\beta}
  \int_{-\Delta\beta}^{\Delta\beta}
  x^2
  \left(
    1-\frac{|x|}{\Delta\beta}
  \right)
  \dd x \\
  &=
  \frac{2}{\Delta\beta}
  \int_{0}^{\Delta\beta}
  x^2
  \left(
    1-\frac{x}{\Delta\beta}
  \right)
  \dd x \\
  &=
  \frac{\Delta\beta^2}{6}.
\end{align}
Thus, the moment-matched covariance is
\begin{equation}
  \Sigma_{s}
  =
  \frac{\Delta\beta^2}{6}\,\bbone.
\end{equation}

\section{Posterior metric under the Fisher Graph Laplacian prior}
\label{sec:diagonalizing_fisher}

In general, the Laplacian matrix is a representation of a graph given by the standard relation
\begin{equation}\label{eq:graph_laplacian}
  L = D - W\, ,
\end{equation} 
where $W$ is the adjacency (or weight) matrix and $D$ is the degree (or degeneracy) matrix. 
The weight matrix is built from the off-diagonal elements of the Fisher matrix
\begin{equation}
  W_{pq} = 
  \begin{cases} 
    F_{pq}, & p \neq q \, , \\ 
    0, & p = q \, . 
  \end{cases}
\end{equation}
By definition, the elements of the degree matrix $D$ sum the rows of $W$, which we can write in terms of the matrix-vector product $F\bm{1}$ 
\begin{equation}
  D_{pp} = \sum_{k \neq p} F_{pk} = (F\bm{1})_p - F_{pp} \, ,
\end{equation}
where $\bm{1}$ correspond to the all-ones vector.
The Laplacian matrix thus becomes
\begin{equation}
  L = \mathrm{diag}(F\textbf{1}) - F\, .
\end{equation}
Substituting $L$ into the posterior metric $G = F + \lambda L$, we get
\begin{equation}\label{eq:posterior_metric}
  G = \lambda\, \mathrm{diag}(F \textbf{1}) + (1 - \lambda) F\, .
\end{equation}

This form makes explicit why the graph Laplacian preserves the row-sums of the Fisher metric. 
For each parameter $p$, the corresponding element of the vector $(F\bm{1})_p = F_{pp} + \sum_{q \neq p}F_{pq}$ 
contains a measure of the total degeneracy of the parameter with the rest of the model. 
A Laplacian matrix always has vanishing row-sums, 
such that adding $\lambda L$ to the Fisher matrix reshapes the degeneracies between parameters without changing the total measure of it.
In the special case $\lambda = 1$, the metric G becomes diagonal and the degeneracy between parameters vanishes.
Nevertheless, the information geometry remains encoded by the total degeneracy measure on the diagonal of the metric.

\section{Matrix-free operators}\label{sec:mf}

This section describes how the action of the Fisher matrix, the graph Laplacian and the diagonal preconditoner 
are evaluated without explicitly materializing them as matrices.

\subsection{Fisher matrix}\label{sec:mff}

Let $J$ denote the Jacobian of the forward model with respect to a set of nuisance parameters.
The corresponding Fisher matrix is
\begin{equation}
  F
  =
  J^\top C^{-1}J\, ,
\end{equation}
where $C$ is the covariance of the noise model.
Since we only require matrix-vector products, we can evaluate the action of the Fisher matrix on a vector $\bm{v}$ as
\begin{equation}
  F\bm v
  =
  J^\top (C^{-1}J\bm v) ,
\end{equation}
where $\bm{u} = C^{-1}J \bm{v}$ can be computed via the Jacobian-vector product and $J^{\top} \bm{u}$ 
via the vector-Jacobian product implemented in \texttt{PyTorch} \citep{torch}.

\subsection{Graph Laplacian matrix}

As in Appendix~\ref{sec:diagonalizing_fisher}, we define the Laplacian matrix by
\begin{equation}\label{eq:graph_laplacian_mvp}
  L = \operatorname{diag}(F\bm 1)-F.
\end{equation} 
Thus, for any vector $\bm v$, we have
\begin{equation}
  L\bm v
  =
  (F\bm 1)\odot \bm v
  -
  F\bm v ,
\end{equation}
where $\odot$ is the Hadamard product.
Since the vector $F\bm 1$ can be computed once and reused, evaluating this operator requires only one matrix-vector product,
which can be computed using automatic differentiation as in the previous section.

\subsection{Preconditioning matrix}
\label{sec:preconditioner}

The posterior metric of the source model takes the general form
\begin{equation}
  G = F + \lambda L \, ,
\end{equation}
This matrix enters the normal equation, described in Section~\ref{sec:matrix-free}, such that a solution to the least-squares minimization 
problem only exists if it is well behaved.
Even with a prior, the wide range in sensitivity between highly magnified source pixels near the caustics 
and unconstrained pixels can slow down the convergence of the conjugate-gradient solver, 
necessitating a preconditioner matrix $M = \operatorname{diag}(G)$ to homogenize its eigenspectrum.

When $\lambda = 1$, evaluating the diagonal of the normal operator simplifies dramatically since $G = \mathrm{diag}(F \textbf{1})$ is already diagonal.
The row-sum vector $F \textbf{1}$ can be evaluated with a matrix-vector product, as described above.
For $\lambda \neq 1$, the preconditioner retains a residual dependency on $\operatorname{diag}(F)$. 
Since computing this vector can be prohibitive, we approximate it with the Hutchinson estimator \citep{Hutchinson1990}. 
We sample a set of $N$ random Rademacher vectors $\bm{z}_i$, whose entries are sampled independently from $\mathcal{U}(\{-1, +1\})$. 
Since $\mathbb{E}[\bm{z}_i \odot F\bm{z}_i] = \operatorname{diag}(F)$, we construct the stochastic estimator
\begin{equation}
  \widehat{\operatorname{diag}}(F) = \frac{1}{N} \sum_{i=1}^{N} \bm{z}_i \odot F\bm{z}_i \,,
\end{equation}
where $\odot$ denotes the Hadamard product.

Each vector requires only a single matrix-vector product $F\bm{z}_i$, which is evaluated efficiently using automatic differentiation.
The general diagonal preconditioner is then assembled element-wise as
\begin{equation}
  M = \operatorname{diag} \left( (1 - \lambda)\widehat{\operatorname{diag}}(F) + \lambda (F\bm{1}) + \epsilon \right) ,
\end{equation}
where $\epsilon$ is a small positive numerical floor used to prevent division by zero in unconstrained parameter directions.

\section{Dark matter population models}\label{sec:population_model}

In the main text, substructure is represented by perturbations of the lensing potential expanded in a spectral basis for the annulus.
This appendix describes how we generate physically motivated realizations of these perturbations from a minimal subhalo population model.

\subsection{Subhalo mass function}

We use a phenomenological model for a population of cold dark matter (CDM) subhalos defined commonly used in the literature 
\citep[e.g.][]{Vegetti2014,Xu2015}
\begin{equation}\label{eq:cdm_smf}
  \frac{\dd n_{\mathrm{CDM}}}{\dd m}  = A_0 \left( \frac{m}{M_{\odot}} \right)^{-\alpha}\,
\end{equation} 
defined as the average number of subhalos per unit of projected area and per unit mass $m$.
This power law is controlled by an amplitude $A_0$ and a logarithmic slope $\alpha = 1.9$.
The mass function is truncated to a mass interval
\begin{equation}
  m_{\min} \leq m \leq m_{\mathrm{max}}\, ,
\end{equation} 
chosen to represent an unresolved population of low-mass subhalos.
We chose $m_{\max} = 10^{9}\, M_{\odot}$ and $m_{\mathrm{min}} = 10^{6}\, M_{\odot}$.

To specify the normalization, we use
\begin{equation}\label{eq:smf_normalization}
A_0 = \Sigma_{\mathrm{crit}} \bar{\kappa}_{\mathrm{sub}}
  \left(
    \int_{m_{\min}}^{m_{\max}}
    m
    \frac{\dd n}{\dd m}
    \,\dd m
  \right)^{-1}
\end{equation} 
where \(\Sigma_{\rm crit}\) is the critical surface density. 
The mean convergence is set to $\bar{\kappa}_{\mathrm{sub}} = 2 \times 10^{-3}$, 
corresponding to 580 subhalos found within the field of view, on average.

\subsection{Warm dark matter model}

To represent warm dark datter (WDM) models, we include a suppression of small-scale structures in the CDM mass function
\begin{equation}
  \frac{\dd n_{\mathrm{WDM}}}{\dd m} = \frac{\dd n_{\mathrm{CDM}}}{\dd m} \left(1 + \frac{m_{\mathrm{hm}}}{m} \right)^{-\beta}\, .
\end{equation} 
The half-mode mass is a phenomenological parameter suppressing the formation of subhalos with masses below the $m_hm = 9 \times 10^{7}\, M_{\odot}$,
and $\beta=1.3$ controls the sharpness of the suppression.
The normalization of this mass function is anchored to the amplitude of the CDM mass function for high mass subhalos, 
which results in 17 subhalos found within the field of view, on average.

\subsection{Sampling the subhalo population}

\begin{figure}[htb!]
  \centering
  \begin{tikzpicture}
    \node at (0, 0) {\includegraphics[width=0.8\textwidth]{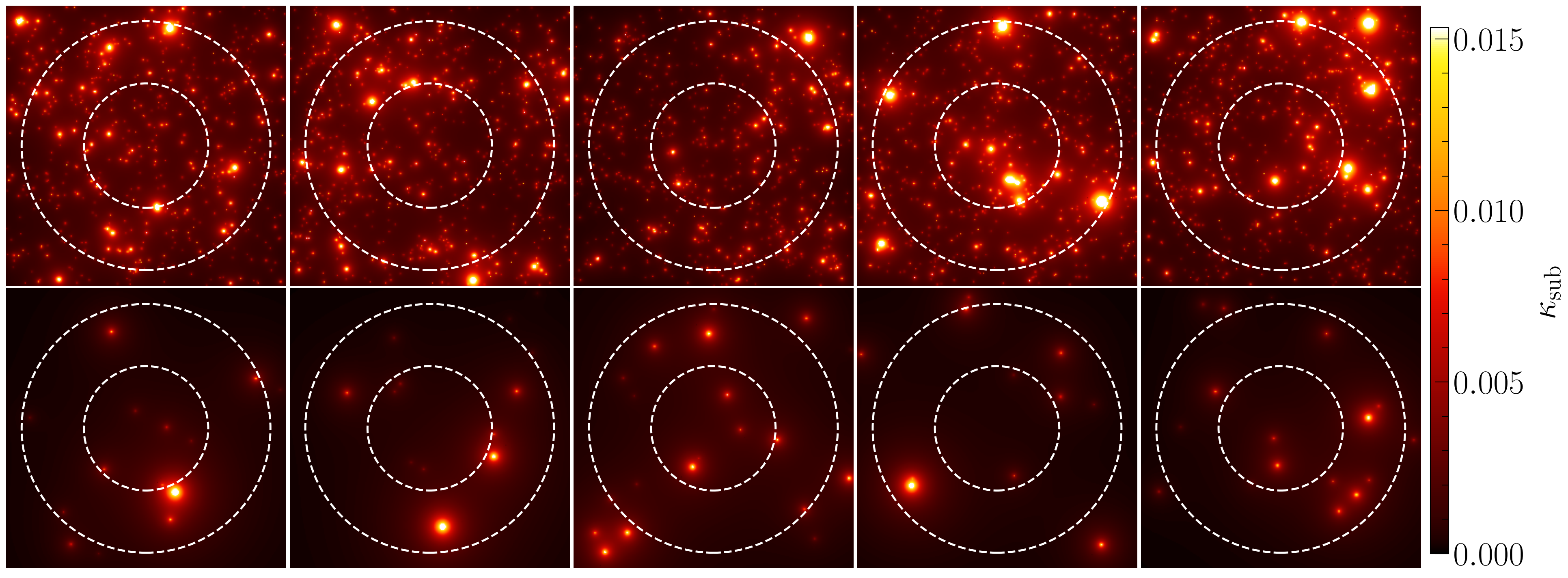}};
    \node[rotate=90, align=center] at (-7.5, 1.25) {CDM};
    \node[rotate=90, align=center] at (-7.5, -1.25) {WDM};
  \end{tikzpicture}
  \caption{Realizations of substructure drawn in a square region with a field of view of $4.5''$, 
  for a cold dark matter (CDM) and a warm dark matter (WDM) model with $m_{\mathrm{hm}} = 9\times 10^{7}\, M_{\odot}$.
  The annulus is only shown for reference, as the convergence has not yet been projected to the eigenbasis.
}
  \label{fig:samples_pop}
\end{figure}

A realization of the population is generated as a Poisson point process.
In practice, the mass interval is separated into 40 bins uniformly spaced in $\log m$. 
In each bin, the subhalo are sampled from a Poisson process
\begin{equation}
  N_{i} \sim \mathrm{Poisson}\left( A_{\mathcal{R}} \int_{m_{i}}^{m_{i + 1}} \frac{\dd n}{\dd m}\, \dd m \right)\, ,
\end{equation} 
where $A_{\mathcal{R}}$ is the area of a square region surrounding the annulus, as in Figure~\ref{fig:samples_pop}.

The location $\bm{\theta}_i$ of each halo is drawn from a uniform spatial distribution in $\mathcal{R}$.
In a more realistic model, the probability of finding subhalos at a specific location 
could depend on the formation history of the system, as well as feedback processes.
The concentration of the NFW profile is related to its mass by \citep[e.g.][]{Gilman2020}
\begin{equation}
    c(m, m_{\rm hm})
    =
    6
    \left(
        \frac{m}{10^{12}M_\odot}
    \right)^{-0.098}
    \left(
        1 + 60\frac{m_{\rm hm}}{m_{200}}
    \right)^{-0.17}.
    \label{eq:gilman18_concentration}
\end{equation}
Given the sampled masses, positions, and concentration,
the substructure convergence field is constructed as a sum of NFW profiles
\begin{equation}
    \kappa_{\rm sub}(\bm{\theta})
    =
    \sum_{i=1}^{N_{\rm tot}}
    \kappa_h(\bm{\theta}-\bm{\theta}_i; m_i, c_i)\, .
    \label{eq:substructure_convergence_sum}
\end{equation}
We do not need to subtract the mean convergence from this field since the eigenfunctions of the annulus eigenbasis are already orthogonal to the constant fonction.
Realizations of substructure are shown in Figure~\ref{fig:samples_pop}.

\end{document}